\documentclass[usenatbib]{mn2e}

\usepackage{txfonts}
\usepackage{graphicx}
\usepackage{natbib}

\def\nat{Nature}
\def\apj{ApJ}
\def\mnras{MNRAS}

\def\araa{ARA\&A}                
\def\aap{A\&A}                   
\def\aaps{A\&AS}                 
\def\aapr{A\&A Rev}          
\def\apjs{ApJS}                  
\def\pasp{PASP}                  
\def\apjl{ApJ}                   
\def\pasj{PASJ}
\def\aj{AJ}
\def\Msun{M$_{\odot}$}
\def\Zsun{$Z_{\odot}$}

\begin{document}

\author[E.~Kuiper et al.]{E.~Kuiper,$^1$\thanks{E--mail: kuiper@strw.leidenuniv.nl} N.~A.~Hatch,$^{1,2}$ H.~J.~A.~R\"{o}ttgering,$^1$ G.~K.~Miley,$^1$  R.~A.~Overzier,$^3$ 
\newauthor B.~P.~Venemans,$^{4}$ C.~De~Breuck,$^4$  S.~Croft,$^{5,6}$ M.~Kajisawa,$^7$ T.~Kodama,$^7$ J.~D.~Kurk,$^8$
\newauthor  L.~Pentericci,$^9$ S.~A.~Stanford,$^{10,11}$ I.~Tanaka,$^{12}$ and A.~W.~Zirm$^{13}$ \\
$^1$ Leiden Observatory, University of Leiden, P.B. 9513, Leiden 2300 RA, the Netherlands \\
$^2$ School of Physics and Astronomy, The University of Nottingham, University Park, Nottingham NG7 2RD \\
$^3$ Max--Planck--Institut f\"{u}r Astrophysik, Karl--Schwarzschild Strasse 1, D--85741 Garching, Germany \\
$^4$ European Southern Observatory, Karl--Schwarzschild Strasse, 85748 Garching bei M\"{unchen}, Germany \\
$^5$ University of California, Merced, PO Box 2039, Merced, CA95344, USA\\
$^6$ University of California, Berkeley, CA94720, USA \\
$^7$ National Observatory of Japan, Mitaka, Tokyo 181--8588, Japan \\
$^8$ Max--Planck--Institut f\"{u}r Extraterrestrische Physik, Giessenbachstrasse, D-85741 Garching, Germany \\
$^9$ INAF, Osservatorio Astronomica di Roma, Via Frascati 33, 00040 Monteporzio, Italy \\
$^{10}$ Institute of Geophysics and Planet Physics, Lawrence Livermore National Laboratory, Livermore, CA 94551, USA\\
$^{11}$ University of California, Davis CA 95616, USA \\
$^{12}$ Subaru Telescope, National Astronomical Observatory of Japan, 650 North Aohoku Place, Hilo, HI 96720, USA \\
$^{13}$ Dark Cosmology Centre, Niels Bohr Institute, University of Copenhagen, Juliane Maries Vej 30, DK-2100 Copenhagen, Denmark \\
}
\title[A census of the galaxies in a protocluster at z$\sim$3]{A galaxy populations study of a radio--selected protocluster at $z\sim3.1$}

\maketitle

\begin{abstract}
We present a population study of several types of galaxies within the protocluster surrounding the radio galaxy MRC~0316--257 at $z\sim3.1$. In addition to the known population of Ly$\alpha$ emitters (LAEs) and [O{\sc iii}] emitters, we use colour selection techniques to identify protocluster candidates that are Lyman break galaxies (LBG) and Balmer break galaxies (BBGs). The radio galaxy field contains an excess of LBG candidates, with a surface density 1.6$\pm0.3$ times larger than found for comparable blank fields. This surface overdensity corresponds to an LBG volume overdensity of $\sim 8\pm4$. The BBG photometric redshift distribution peaks at the protocluster's redshift, but we detect no significant surface overdensity of BBG. This is not surprising because a volume overdensity similar to the LBGs would have resulted in a surface density of $\sim1.2$ that found in the blank field. This could not have been detected in our sample. Masses and star formation rates of the candidate protocluster galaxies are determined using SED fitting. These properties are not significantly different from those of field galaxies.
The galaxies with the highest masses and star formation rates are located near the radio galaxy, indicating that the protocluster environment influences galaxy evolution at $z\sim3$. We conclude that the protocluster around MRC~0316--257 is still in the early stages of formation.
\end{abstract}
\begin{keywords}
galaxies: evolution -- galaxies: high-redshift -- galaxies: clusters: general -- cosmology: observations -- cosmology: early Universe
\end{keywords}

\section{Introduction} \label{sec:intro}

One of the main aims of astrophysics is to understand the formation and evolution of galaxies.  Hierarchical evolution in $\Lambda$CDM cosmology means the evolution of a galaxy will depend on whether it is located in a low-density or high-density environment \citep{toomre1977}. This has been quantified in studies by \citet{clemens2006} and \citet{sanchez2006} for the local Universe and for the more distant Universe by \citet{vandokkum2007} and \citet{gobat2008}. These studies find that early-type galaxies in cluster environments are older than early-type galaxies residing in low-density, field environments, indicating that galaxies in cluster-like environments form at an earlier epoch. 

Further evidence that environment influences galaxy evolution is the observation of `environment-dependent downsizing' in galaxy clusters at $z\sim1$. Downsizing  \citep{cowie1996} implies that the bright, massive galaxies move onto the red sequence first, whilst fainter galaxies are added at a later time. \citet{tanaka2005,tanaka2007,tanaka2008} show that the red sequence in high density environments extends to fainter magnitudes than in less dense galaxy groups. The red sequence therefore forms at an earlier epoch in dense environments. 

To understand galaxy evolution in different environments it is necessary to study galaxy clusters across cosmic time. Galaxy clusters have been detected out to $z=1.5$ using conventional techniques such as through observation of the hot X--ray emitting intra-cluster gas or IR red sequence searches \citep[e.g.,][]{mullis2005,stanford2005,stanford2006}. 

The most successful technique for finding cluster progenitors at $z>1.5$ is to search for emission line galaxies around high-redshift radio galaxies \citep[HzRGs, for a comprehensive review see][]{miley2008}. HzRGs are among the most luminous and massive objects in the early Universe and are expected to be the progenitors to local cD galaxies \citep{roccavolmerange2004,seymour2007}. Several studies have found that they are situated in overdense regions with properties expected of forming galaxy clusters \citep{pentericci2000,venemans2005,venemans2007,intema2006,overzier2006,overzier2008}. The dense cluster-like environments around HzRGs are likely not yet virialized and are commonly termed `protoclusters'. They are excellent laboratories for studying the formation and evolution of galaxies in overdense environments. 

In this study we investigate galaxy populations around the HzRG MRC~0316--257 (hereafter 0316) located at $z=3.13$. Galaxy selection techniques based on the Lyman break at 912~\AA~\citep[e.g.,][]{steidel2003} are most efficient at this redshift. Also at this redshift, strong emission lines fall within existing narrowband filters. This allows the selection of many different types of emission line galaxies in the protocluster. Therefore the redshift of this protocluster makes it ideal for studying its galaxy populations. \citet{venemans2005} (hereafter V05) showed this region contains an overdensity of Ly$\alpha$ emitters (LAEs) and \citet{maschietto2008} (hereafter M08) found a number of [O{\sc iii}] emitters near the redshift of the radio galaxy (RG). Additional galaxy populations are identified using a large set of broadband images. The properties of the galaxy populatons are determined using broadband photometry and spectral energy distribution (SED) fitting. These properties are then compared to the properties of field galaxies to search for environmental influences on galaxy evolution at this redshift.

The paper is ordered as follows: in Sect.~\ref{sec:data} the data and its reduction is discussed. This is followed by the photometry in Sect.~\ref{sec:photom} and sample selection in Sect.~\ref{sec:sample}. The particulars of the SED fitting process are discussed in Sect.~\ref{sec:fitting}, after which we show our results in Sect.~\ref{sec:res}. We discuss and compare the results to work by other authors in Sect.~\ref{sec:disc}. Finally, the summary and conclusions are presented in Sect.~\ref{sec:conc}. Throughout this paper we use a standard $\Lambda$CDM cosmology with $H_0$=71~km~s$^{-1}$~Mpc$^{-1}$, $\Omega_{\rm M}$=0.27 and $\Omega_{\Lambda}$=0.73.  All magnitudes given in this paper are in the AB magnitude system \citep{oke1983} unless noted otherwise.

\section{Data} \label{sec:data}

\begin{figure*}
\resizebox{\hsize}{!}{\includegraphics{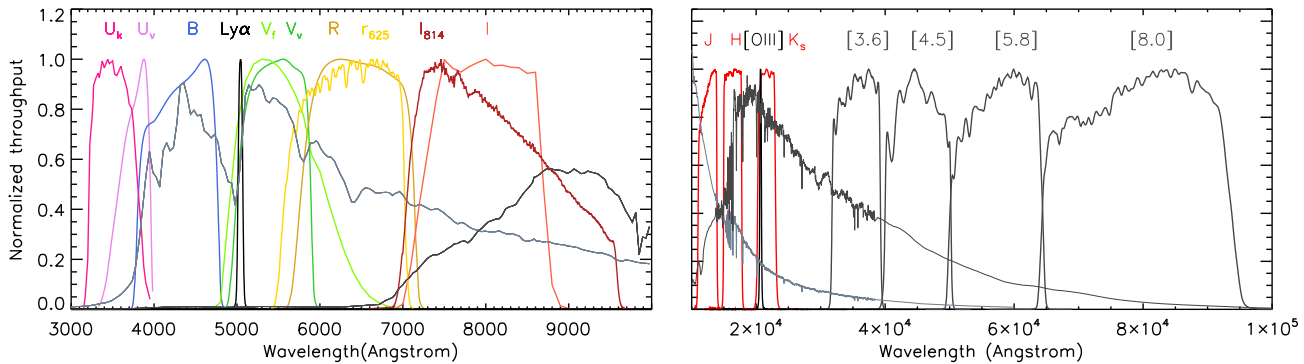}}
\caption{\label{fig:rcurves} The filter response curves for the filters used in this study. For illustrative purposes the curves have been scaled to the same maximum throughput. The $J_{\rm m}$ and $K_{\rm m}$ filter curves are not shown due to their similarity to the corresponding ISAAC filters. Also plotted are two $z=3.13$ model SEDs obtained from the BC03 population synthesis models. The light gray SED is for a continuous star forming galaxy of 100~Myr old, whereas the dark gray SED is for a 1~Gyr old galaxy which has an exponentially declining star formation with $\tau= 10$~Myr.}
\end{figure*}

Images of the 0316 field were obtained in 20 passbands spanning the $U$ band to the 8~$\mu$m band. A summary of all the data used is given in Table 1 and the filter response curves of each of the filters is shown in Fig.~\ref{fig:rcurves} together with two example SEDs taken from the \citet{bruzual2003} (BC03) models. The fields covered by the various instruments are illustrated in Fig.~\ref{fig:allobs}. The reduction of the various data sets is described below.

\subsection{Ground-based UV-optical imaging}

$UBVR$ imaging data were obtained using the VIsible MultiObject Spectrograph \citep[VIMOS,][]{lefevre2003} instrument at the Very Large Telescope (VLT). The data were taken during the period of 14--15 November 2003 for the V band and 20--25 November 2003 for the remaining bands. Additional $U$ band data were obtained on 14 February 2004. 

The VIMOS field-of-view consists of 4 separate quadrants, each having a field-of-view of approximately 7'$\times$8'. Two pointings were used in which the RG was centered in one of the quadrants. For most of this work only the central quadrant that contains the RG is used, as only this central region has additional data.

\begin{table*}
\caption{\label{table} Details of the observations. The 5$\sigma$ limiting magnitudes have been calculated for angular diameters of 2\arcsec\ for the ground based data, 0.5\arcsec\ for the ACS data and 4\arcsec\ for the IRAC data. }
\begin{tabular}{c|c|c|c|c|c|c|c}
\hline
Band & Instrument & Date of observation & $\lambda_{\rm eff}$(\AA) & $\Delta\lambda$(\AA) & Exp. time (sec.) & Seeing & 5$\sigma$ limiting mag.\\
\hline
\hline
$U_{\rm k}$ & LRIS/Keck & 2003 Jan. 31, 2003 Feb. 1 \& 4  & 3516 & 561 & 17100 & 1.3\arcsec & 26.33 \\
\hline
$U_{\rm v}$ & VIMOS/VLT & 2003 Nov. 23--24, 2004 Feb. 14 & 3744 & 359 & 15500 & 1.0\arcsec & 25.98 \\
$B$ & VIMOS/VLT & 2003 Nov. 22  & 4310 & 832 & 3720 & 0.8\arcsec & 26.02 \\
$V_{\rm v}$ & VIMOS/VLT & 2003 Nov. 14--15 & 5448 & 842 & 4960 & 0.75\arcsec & 25.97 \\
$V_{\rm f}$ & FORS2/VLT & 2001 Sept. 20--21 & 5542 & 1106 & 4860 & 0.7\arcsec & 26.10 \\
$R$ & VIMOS/VLT & 2003 Nov. 20 \& 25 & 6448 & 1292 & 6845 & 0.9\arcsec & 25.90 \\
$I$ & FORS2/VLT & 2001 Sept. 6--8 & 7966 & 1433 & 4680 & 0.65\arcsec & 25.83 \\
\hline
$r_{625}$ & ACS/HST & 2004 Dec. 14--31, 2005 Jan. 2--21& 6321 & 1327 &  23010 & - & 27.66  \\
$I_{814}$ & ACS/HST & 2002 Jul. 18, 2004 Dec. 14--31, 2005 Jan. 2--21  & 8089 & 1765 & 52320 & - & 28.37 \\
\hline
$J_{\rm i}$ & ISAAC/VLT & 2003 Nov. -- 2004 Oct. & 12535 & 2640 & 19000 & 0.5\arcsec & 24.60 \\
$K_{\rm i}$ & ISAAC/VLT & 2003 Nov.-- 2004 Oct. & 21612 & 2735 & 19000 & 0.5\arcsec & 24.19\\
\hline
$J_{\rm m}$ & MOIRCS/Subaru & 2006 Jan. 6--7 & 12532 & 1538 & 4680 & 0.75\arcsec & 23.77 \\
$H$ & MOIRCS/Subaru & 2006 Jan. 6--7 & 16364 & 2788 & 3600 & 0.8\arcsec & 22.71\\
$K_{\rm m}$ & MOIRCS/Subaru & 2006 Jan. 6--7 & 21453 & 3042 & 3300 & 0.75\arcsec & 23.07 \\
\hline
[3.6] & IRAC/Spitzer & 2005 Jan. 17--19 & 35636 & 6852 & 46000 & - & 23.76 \\

[4.5] & IRAC/Spitzer & 2005 Jan. 17--19 &  45111& 8710 & 46000 & - & 23.60 \\

[5.8] & IRAC/Spitzer & 2005 Jan. 17--19 & 57598 & 12457 & 46000 & - & 22.24 \\

[8.0] & IRAC/Spitzer & 2005 Jan. 17--19 & 79594 & 25647 & 46000 & - & 21.97 \\
\hline
Ly$\alpha$ NB & FORS2/VLT & 2001 Sept. 20--21 & 5040 & 61 & 23400 & 0.7\arcsec & 25.3\\

[O{\sc iii}] NB & ISAAC/VLT & 2003 Nov. -- 2004 Oct. & 20675 & 437 & 24840 & 0.45\arcsec & 22.6 \\
\hline
\end{tabular}
\end{table*}

The reduction of the UV and optical data was performed using standard tasks in the IRAF\footnote[13]{IRAF is distributed by the National Optical Astronomy Observatory, which is operated by the Association of Universities for Research in Astronomy, Inc., under cooperative agreement with the National Science Foundation.} software package. The process includes bias subtraction and flat fielding using twilight sky flats. Remaining large scale gradients were removed using a smoothed master flat. This was obtained by median-combining the unregistered, flatfielded science images. Reduced images were registered and combined to form the final science images. Photometric zeropoints were determined using standard star images taken on the same nights as the science frames. 

Deep Keck $u'$ band data ($U_{\rm k}$)  were obtained on 31 January, 1 February and 4 February 2003 using the blue arm of the Low Resolution Imaging Spectrometer \citep[LRIS,][]{oke1995}. As there are large differences in the filter responses, both $U$ bands have been included in the analysis. Information regarding the reduction of these data can be found in \citet{venemans2007}.

$V$ and $I$ band data of the 0316 field were obtained during the respective periods of 20--21 and 6--8 September 2001 using the FORS2 instrument at VLT. Information on these data and their reduction can be found in V05.

\subsection{HST/ACS optical imaging}

Deep Hubble Space Telescope (HST) $r_{625}$ and $I_{814}$ images of the 0316 field cover approximately half of the VIMOS field-of-view. These images were obtained using the Advanced Camera for Surveys \citep[ACS,][]{ford1998} during the periods of 14--31 December 2004 and 2--21 January 2005. The data covered two fields of 3.4\arcmin$\times$3.4\arcmin~with approximately 1\arcmin~overlap between the two fields. The two $I_{814}$ fields were combined with an additional 3.4\arcmin$\times$3.4\arcmin~ACS field which was obtained on 18 July 2002. Details of the data and their reduction can be found in M08 and V05.

\subsection{Near-infrared data}

Two sets of near-infrared (NIR) data spanning the $J$ to $K_{\rm s}$ bands were used. Deep $J$ and $K_{\rm s}$ images were obtained with the Infrared Spectrometer And Array Camera \citep[ISAAC,][]{moorwood1998} on the VLT on various dates between November 2003 and October 2004 (see M08). These images are deep, but only cover the innermost 2.5\arcmin$\times$2.5\arcmin~of the protocluster.

Additional $JHK_{\rm s}$ data were obtained using the Multi-Object InfraRed Camera and Spectrograph \citep[MOIRCS,][]{ichikawa2006,suzuki2008} at the Subaru telescope on 6--7 January 2006. For details concerning the reduction of these data we refer the reader to \citet{kodama2007}, hereafter K07. This set of $JHK_{\rm s}$ imaging data is shallower than the ISAAC data described above, but it covers a larger fraction of the 0316 field. To avoid confusion between the ISAAC and MOIRCS data we denote the bands with a subscript `i' or `m', respectively.

\subsection{Mid-infrared data}

In the mid-infrared wavelength range we have Spitzer InfraRed Array Camera \citep[IRAC,][]{fazio2004} data at 3.6~$\mu$m, 4.5~$\mu$m, 5.8~$\mu$m  and 8.0~$\mu$m (hereafter [3.6], [4.5], [5.8] and [8.0] respectively). IRAC data were obtained in all bands on 17--19 January 2005 covering a $\sim$5\arcmin$\times$5\arcmin~field centred on the RG. Deep imaging was obtained using a medium scale cycling dither pattern of 230 frames with a 200~s frame time for a total exposure time of $\sim$12.7  hours. The [3.6] and [4.5] basic calibrated level data (BCD) were reduced and mosaiced using the {\sc mopex} software \citep{makovoz2005} following standard procedures. Before the BCD frames were combined, the muxbleed and column pulldown effects were corrected using custom software provided by D.\ Stern and L.\ Moustakas. During mosaicing, the images were resampled by a factor of $\sqrt2$ and rotated by 45$^{\circ}$. The [5.8] and [8.0] BCD data were further mosaiced using a custom IDL code kindly provided to us by I. Labb\'{e}.

\subsection{Further reduction}

With the exception of the ACS data, all images were resampled to a common pixel scale of 0.205\arcsec/pixel and transformed to the same image coordinate system using the IRAF tasks {\sc geomap} and {\sc gregister}. The images were then convolved with 2D Gaussian profiles to match the PSF FWHM of the VIMOS $U$  band ($U_{\rm v}$), which has the largest seeing of approximately 1\arcsec. The IRAC bands and the $U_{\rm k}$ band are excluded from this process as they have significantly larger PSF sizes. Smoothing the other images to the PSF size of these images would negatively impact the quality of the analysis.

Because of the extreme smoothing required to match the resolution of the ACS images to that of the ground-based data, the ACS data were not used for the SED fitting and photometric redshift determination, but only for the determination of UV slopes of the protocluster galaxy candidates.

\section{Photometry} \label{sec:photom}

Photometry was obtained using the {\sc SExtractor} software \citep{bertin1996} in double image mode. Lyman Break Galaxy candidates (LBGs) were detected using the unsmoothed $R$ band as the detection image. For the Balmer Break Galaxy candidates (BBGs) the unsmoothed $K_{\rm s}$ bands were used as detection images. 

A detection is defined as an object with 5 adjacent pixels that each exceed the 3$\sigma$ rms noise. Colours were measured in the 3$\sigma$ level isophot apertures as determined from the detection image. If a known object was undetected in an image it was assigned the 3$\sigma$ detection limit of the band in question. Lowering this limit to $1\sigma$ does not affect any of the conclusions presented in this paper. Total magnitudes and fluxes in the detection bands were obtained by using the MAG\_AUTO function of {\sc SExtractor}. Total fluxes in the remaining bands were obtained by scaling the respective isophotal magnitudes accordingly. 

Because of the large size (FWHM) of the PSF in the IRAC images it was not possible to determine consistent colours using the isophotal apertures mentioned above. Instead, the object flux was measured in a circular aperture of 20 pixel (4\arcsec), and a correction factor applied to match the photometry to the other bands. This correction factor was determined by smoothing the detection images to the spatial resolution of the IRAC data. Object fluxes were measured in the smoothed detection images in 4\arcsec\ circular apertures, and compared to the flux measured from the unsmoothed detection images in the standard 3$\sigma$ isophotal apertures. The ratio between the two fluxes yielded the correction factor. A similar process was applied to determine the colours in the $U_{\rm k}$ image. 

The large FWHM of the PSF in the IRAC bands also causes source confusion and contamination by neighbouring sources. For faint objects this effect can be a large source of error. All detected objects were visually inspected for contamination, and heavily contaminated objects have been removed from the analysis when relevant.

Finally, all magnitudes were corrected for Galactic foreground extinction determined from the \citet{schlegel1998} extinction maps.

Photometric uncertainties and limiting magnitudes (listed in Table~\ref{table}) were computed using the method described in \citet{labbe2003}, which is summarized below. After masking all objects, the rms of pixels in the entire image and the rms of fluxes in apertures of various sizes were measured. We determined the noise in an aperture of size $N$ using the relation
\begin{equation}
\sigma_{\rm i}(N)=N\bar{\sigma_{\rm i}}(a_{\rm i}+b_{\rm i}N)
\end{equation}
with $\bar{\sigma_{\rm i}}$ the rms of pixels over the entire image and $\sigma_{\rm i}$ the rms within a certain aperture size $N$. $N$ is defined as $N=\sqrt{A}$ where $A$ is the area of the aperture. The subscript `i' indicates the photometric band in question. The free parameters $a_{\rm i}$ and $b_{\rm i}$ were then fitted such that it can be calculated what the noise is in an aperture of a given size. The uncertainty calculated using the rms of all the background pixels can be an underestimate, because it does not take into account pixel-to-pixel dependencies introduced in the reduction of the data. Due to the IRAC photometry being more uncertain an additional 10 per cent uncertainty was added in quadrature for those four bands as has been done in previous studies \citep[e.g.][]{labbe2005}. 

\begin{figure}
\resizebox{\hsize}{!}{\includegraphics{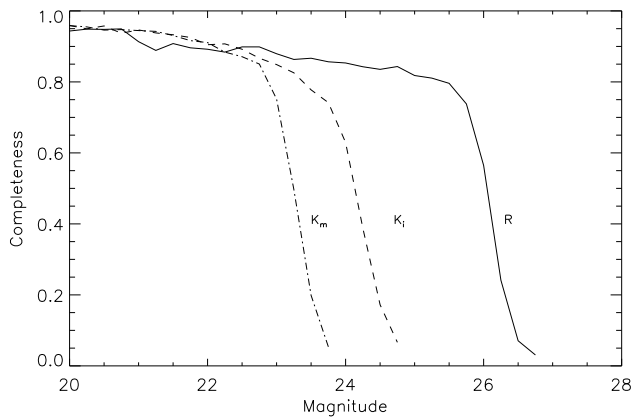}}
\caption{\label{fig:compl}Completeness of the three detection bands as a function of magnitude for points sources.}
\end{figure}

The completeness of the detection fields for point sources is shown in Fig.~\ref{fig:compl}. The completeness was determined by extracting several bright, unsaturated stars from the images, averaging them to obtain a high signal-to-noise PSF image and then inserting a number of these PSFs at random locations in the images. The source extraction was repeated and the number of recovered artificial stars yielded the completeness as a function of magnitude. To avoid overcrowding only 150 objects were added at a time. This process was repeated ten times for each magnitude to obtain better statistics. Due to the small field size of the ISAAC data the number of added objects was lowered to 50 and the process was repeated fifty times. The data are 50 per cent complete down to $R=26$, $K_{\rm i}=24.1$ and $K_{\rm m}=23.2$.

\section{Sample selection} \label{sec:sample}

\subsection{Ly$\alpha$ and [O{\sc iii}] excess objects} \label{sec:sampex}

\citet{venemans2005} spectroscopically confirmed 32 LAEs at the redshift of the RG, and found that the volume density of LAEs in the 0316 field is 2--4 times larger than for the blank field. 

\citet{maschietto2008} found a sample of 13 [O{\sc iii}] emitting galaxies near the RG, corresponding to a surface density $3.5_{-2.2}^{+5.6}$ times the field density (uncertainties obtained from Poisson statistics). The [O{\sc iii}] overdensity is consistent with the LAE overdensity found by V05, but the sample is small and the uncertainties are large. Five of the [O{\sc iii}] emitters are spectroscopically confirmed LAEs, and three additional [O{\sc iii}] emitters are spectroscopically confirmed to be at $z=3.1$. These [O {\sc iii}] emitters are blueshifted with respect to the RG indicating that the protocluster around 0316 may be part of a larger superstructure. 

\subsection{Lyman Break candidates} \label{sec:samplbg}

LBGs were selected using a colour criterion similar to that used by \citet{steidel2003} to select star forming galaxies at $z\sim 3$. The criterion uses the $U_{\rm k}VR$ bands. Even though the VIMOS field is larger, the $U_{\rm v}$ passband is redder than $U$ passbands used in most other studies. Thus the $U_{\rm k}$ band was used to facilitate comparison with other LBG studies. 

The colour criterion was devised by creating artificial galaxy spectra using the BC03 evolutionary population synthesis models. Galaxy spectra were synthesized for various star formation histories (SFHs) and a variety of values for extinction and age. The spectra were redshifted from $z=0$ to $z=5$. The model spectra were then convolved with the filter curves to obtain synthetic galaxy photometry. Galaxies situated at $3.0 < z < 3.3$ lie in the upper left corner of the $U_{\rm k}-V_{\rm v}$ vs. $V_{\rm v}-R$ colour-colour diagram. This region is parametrized by the relations
\begin{eqnarray} \label{eq:crit}
& U_{\rm k}-V_{\rm v} \ge 1.9, & \nonumber \\
& V_{\rm v}-R \le 0.51,  & \\
& U_{\rm k}-V_{\rm v} \ge 5.07\times (V_{\rm v} -R)+2.43, & \nonumber \\
& R \le 26. & \nonumber
\end{eqnarray}

\begin{figure}
\resizebox{\hsize}{!}{\includegraphics{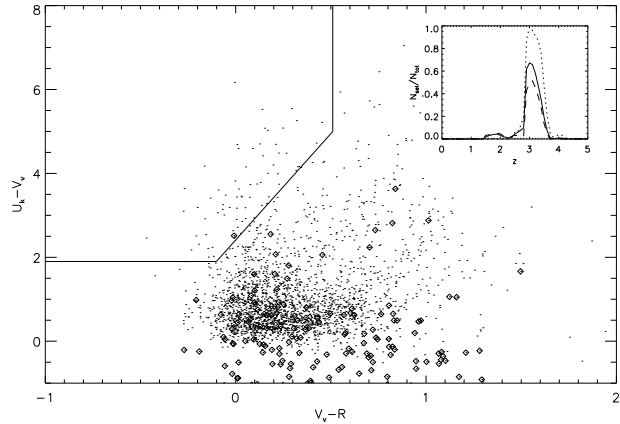}}
\caption{\label{fig:colcol} $U_{\rm k}-V_{\rm v}$ vs. $V_{\rm v}-R$ diagram of all objects detected in the $R$ band. Diamonds indicate the points with lower limits to their $U_{\rm k}-V_{\rm v}$  colours. The lines indicate the area in colour-colour space that was used for the LBG selection. The inset shows the selection efficiency for all objects (full line), objects that are younger than 100~Myr and have $E(B-V) < 0.3$ (dotted line) and objects that are older than 500~Myr (dashed line). The criterion should predominantly select objects in the range $2.9 < z < 3.4$.}
\end{figure}

The $U_{\rm k}-V_{\rm v}$ vs. $V_{\rm v}-R$ colours for the $R$ band detected sample are shown in Fig.~\ref{fig:colcol}. The LBG selection criterion is marked as solid lines. The inset shows the ratio of the number of synthesized objects that are selected to the total number of synthesized objects as a function of redshift. This criterion should predominantly select objects having redshifts between $2.9 < z < 3.4$ which have low ages and little dust obscuration ($t<100$~Myr, $E(B-V) < 0.3$). The 50 per cent completeness limit of $R=26$ was adopted as a magnitude cut. A total of 52 objects in the 0316 field satisfy the selection criterion.

Photometric redshifts ($z_{\rm phot}$) were determined for all $R$ band detected objects using the EAZY code \citep{brammer2008}. Since a significant number of objects lack deep NIR coverage we include the more uncertain IRAC bands. To test the influence of the IRAC data on the photometric redshift determination we have determined the photometric redshifts both including and excluding the IRAC photometry.

The EAZY code yields 4 different estimates of the redshift. It offers the options of applying both marginalization\footnote[14]{The process of marginalization calculates the best photometric redshift value by weighting it according to the redshift probability distribution.} and a Bayesian prior. In order to determine which option is best suited for this study we used the Multiwavelength Survey by Yale--Chile (MUSYC) ECDF-S data \citep{gawiser2006,damen2009,taylor2009} to produce an $R$ band detected catalogue. The photometric redshifts of this sample were determined using EAZY and subsequently compared to a large sample of spectroscopic redshifts ($z_{\rm spec}$) from the catalogues of \citet{cimatti2002}, \citet{lefevre2004} and \citet{ravikumar2007}. A total of 2469 objects have spectroscopic redshifts. Note that this will not be a completely consistent test case for the 0316 data, because the objects for which spectroscopic redshifts are available are in general brighter than $R=25$, and the details concerning the depths in certain MUSYC bands are different from the 0316 dataset. 

\begin{figure}
\resizebox{\hsize}{!}{\includegraphics{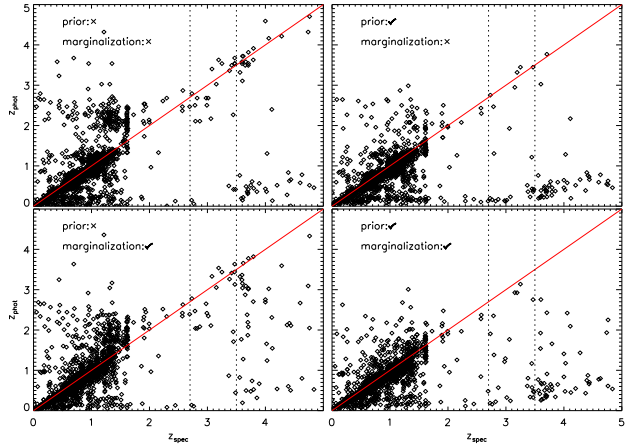}}
\caption{\label{fig:zzzs}$z_{\rm phot}$ versus $z_{\rm spec}$ for a sample of 2469 objects detected in the ECDF-S. No magnitude cut was applied. Photometric redshifts were obtained using the EAZY redshift code and include data in the $UBVRIJHK_{\rm s}$[3.6][4.5] bands. The four panels represent the results for the four different redshift determinations. Each panel indicates whether a prior or marginalization was used. The vertical lines mark the redshift interval of interest. It is apparent that it is best to use neither prior nor marginalization for $z>2$ objects.}
\end{figure}

The comparison between $z_{\rm spec}$ and $z_{\rm phot}$ for the MUSYC galaxies is shown in Fig.~\ref{fig:zzzs}. The four panels show the results for four different methods of obtaining the photometric redshift. The scatter is quantified by the median absolute deviation of $dz=(z_{\rm spec}-z_{\rm phot})/(1+z_{\rm spec})$. For $z_{\rm spec}<$ 1.5 applying both the prior and the marginalization yields the least amount of scatter ($|dz|=0.043$) and gives in general good agreement. However, application of the prior also shifts the objects with $z_{\rm spec} > 2$ to $z_{\rm phot} \sim 0.1$. 

Since we target galaxies at $z\sim3$, the photometric redshift determination without prior or marginalization yields the best results. Comparing the scatter for $z_{\rm spec} > 2$ yields $|dz|=0.22$ for the upper left panel and $|dz|=0.69$ and $|dz|=0.65$ for the upper and lower right panels, respectively. The option that uses neither prior nor marginalization systematically overestimates the redshifts of objects at $z\sim1$, placing them at $2 < z_{\rm phot} < 2.5$ instead. However, only 0.2 per cent of objects with $z_{\rm spec} < 1.5$ are placed in the range $2.7 < z_{\rm phot} < 3.5$, so this should have no significant effect on the results. In the remainder of this paper, photometric redshifts of the 0316 galaxies have been obtained using neither prior nor marginalization.

\begin{figure}
\resizebox{\hsize}{!}{\includegraphics{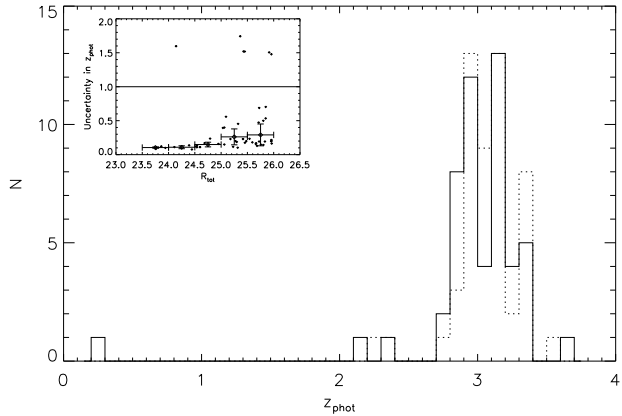}}
\caption{\label{fig:zphot} Redshift distributions of the sources selected using the colour cuts in Eq.~\ref{eq:crit}. The full line and dotted line denote results obtained with and without the IRAC bands, respectively. In the inset the redshift uncertainty $\Delta z$ is plotted as a function of $R$ band magnitude for all object that satisfy the LBG colour selection criterion. Also shown are average values for 0.5 mag bins. Objects that have uncertainties larger than 1 (as indicated by the horizontal line) are not included in these average values. These objects are degenerate between $z \sim 0$ and $z \sim 3$ and would skew the average to higher values.}
\end{figure}

The photometric redshift distribution of the 0316 LBG candidates is shown in the main panel of Fig.~\ref{fig:zphot}. The majority of objects is situated at $z \sim 3.1$ and strong peaks are seen at the redshift of the protocluster and $z\sim 2.9$. These peaks are artefacts caused by the discrete number of possible redshifts. Excluding the IRAC data (dotted line) has little influence on the photometric redshift distribution. This is due to the inclusion of both $U_{\rm k}$ and $U_{\rm v}$. Both bands sample the strong Lyman break feature of these galaxies and hence the photometric redshifts are well constrained. 

All objects within $2.7 < z < 3.5$ were selected for the final sample of 48 objects. This is based on the photometric redshift uncertainties shown in the inset of Fig.~\ref{fig:zphot}. Here the redshift uncertainty $\Delta z$ is taken to be half of the 1$\sigma$ uncertainty interval yielded by EAZY. The inset indicates that an uncertainty of 0.4 is a reasonable value to encompass all possible $z\sim 3.1$ objects. The accuracy of photometric redshifts is not sufficient to verify if any of these objects are indeed members of the protocluster surrounding 0316; to achieve this spectroscopic redshifts are needed. 

\subsection{Balmer Break Galaxy candidates} \label{sec:sampdrg}

The Balmer break can be used to select $z\sim3$ galaxies in a similar fashion to the Lyman break method discussed in Sect.~\ref{sec:samplbg}.

Extracting a sample of distant red galaxies (DRGs) was done by using the simple colour cut $(J-K_{\rm s} )_{\rm Vega}\ge2.3$ proposed by \citet{franx2003}. The DRG colour criterion samples red, predominantly massive galaxies between a redshift of 2 and 4. Although the NIR filter sets used in this study differ from other studies the $(J-K)$ colour difference is typically $\sim$0.02~mag, which is small compared to the photometric errors. Therefore these colour terms are deemed negligible.

Thirty-four DRGs are found in the ISAAC field-of-view after removing 6 ghost images. After applying a 50 per cent completeness cut at $K_{\rm s}=24$ the final sample of 17 DRGs remains. Amongst the DRGs are the RG, an LAE and an [O{\sc iii}] emitter with $z_{\rm spec}=3.104$ (M08).

The MOIRCS data is about 1 magnitude shallower in the $K_{\rm s}$ band than the ISAAC data. K07 apply a magnitude cut of $K_{\rm m}=23.7$ and find a total of 54 DRGs in the MOIRCS field, 14 of which are located in the ISAAC field. Using the detection criterion described above (5 adjoining pixels 3$\sigma$ above the background) leads to several spurious detections in the MOIRCS data compared to the ISAAC data. Increasing the detection criterion to 3.25$\sigma$ removes most of the spurious detections including two DRGs detected by K07. These are spurious detections since neither of these two objects have counterparts in the deeper ISAAC data. There may also be spurious DRGs outside the ISAAC field-of-view. DRGs with $K_{\rm m}>23$ (the 50 per cent completeness level) were removed from the sample resulting in a final sample of 23 DRGs in the MOIRCS field. The difference in the number of DRGs in the K07 study and the present study is due to the stricter magnitude cut and the different parameter values used for source detection. 

Combining the two separate DRG samples leads to a sample of 33 unique DRGs. In the rest of the paper we will refer to DRGs as Balmer Break Galaxies (BBGs).

\begin{figure}
\resizebox{\hsize}{!}{\includegraphics{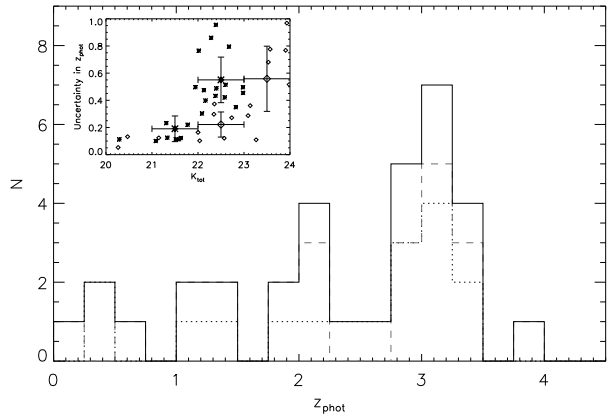}}
\caption{\label{fig:zphotdrg} Redshift distribution of the ISAAC detected BBGs and the MOIRCS detected BBGs are shown as a dotted and dashed line, respectively. The redshift distribution for the combined sample is also shown as the full line. Note the peak in number at the redshift of the radio RG.}
\end{figure}

The main panel of Fig.~\ref{fig:zphotdrg} shows the photometric redshift distribution that is found for both BBG samples and the combined sample. We find that there is a peak at $z \sim 3$, the redshift of the protocluster. The inset shows the uncertainties on the photometric redshifts as a function of $K_{\rm s}$ magnitude.  As can be seen from the inset the uncertainty on the photometric redshift increases rapidly as the magnitude becomes fainter, with 1$\sigma$ uncertainties $>0.5$ for $K_{\rm i} \ge 23$. This is due to the faint $U_{\rm k}$ and $U_{\rm v}$ magnitudes of these objects resulting in a poorly constrained Lyman break. Such large uncertainties would smooth any peak in the redshift distribution caused by the protocluster.

A Kolmogorov--Smirnov test was used to compute the significance of the peak using the photometric redshift distribution of BBGs from \citet{grazian2007} as a comparison sample. There is a probability of 0.3 per cent that both distributions have been drawn from the same parent distribution. Thus the photometric redshift distribution of BBGs in the 0316 region differs at the 3$\sigma$ level from that in the field. 

Possible protocluster members are identified as the BBGs that lie within $2.7 < z_{\rm phot} < 3.5$. A total of 9 ISAAC BBGs satisfy this criterion. For the MOIRCS BBGs a similar fraction of 11 out of 23 lie within $2.7 < z_{\rm phot} < 3.5$. 

\subsection{Spectroscopic redshifts}

\begin{figure}
\resizebox{\hsize}{!}{\includegraphics{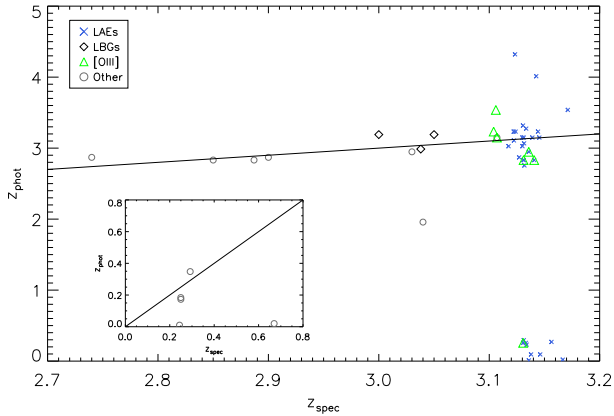}}
\caption{\label{fig:specphot} $z_{\rm phot}$ versus $z_{\rm spec}$ for all the objects in the 0316 field that have spectroscopic redshifts. The line indicates equality. The main panel shows the redshift range of interest, whereas the inset shows low redshift objects. Circles denote the 13 objects that are unclassified, the diamonds are LBGs, the blue crosses denote the LAEs and the green triangles indicate the [O{\sc iii}] emitters. The RG is the [O{\sc iii}] emitter in the lower right corner of the figure.}
\end{figure}

Within the 0316 field spectroscopic redshifts are known for all 32 LAEs, 7 [O{\sc iii}] emitters and 16 additional objects (Bram Venemans, private communication). Fig.~\ref{fig:specphot} compares these spectroscopic redshifts to the photometric redshifts obtained with the EAZY photometric redshift code. 

In general the photometric redshifts agree well with the spectroscopic redshifts. Even the intrinsically faint LAEs (denoted by blue crosses) show good agreement. The scatter around the protocluster's redshift of approximately $\sigma_{z}=0.3$ is similar to the width chosen for the redshift cut in the LBG and BBG selection. Because of this good agreement we expect that the photometric redshift estimates are adequate representations of the true redshifts.

There are six additional objects with $z_{\rm spec}>3$, but only three of these are classified as LBGs. It thus seems that approximately half of the $z\sim3$ galaxies have been missed have by the colour selection technique. We discuss this further in Sect.~\ref{sec:rest}. Furthermore, the photometric redshift of the RG differs strongly from its spectroscopic redshift. This is due to its proximity to a foreground galaxy which affects the photometry.

\subsection{Completeness: photometric redshift selection} \label{sec:rest}

In addition to the selected protocluster galaxy candidates, we selected unclassified objects that have photometric redshifts in the range $2.7 < z_{\rm phot} < 3.5$. The size of this sample gives an estimate of the number of $z\sim 3$ galaxies that are missed by the above-mentioned colour-selection techniques.

Galaxies with $2.7 < z_{\rm phot} < 3.5$ were selected from the $R$ band and $K_{\rm s}$ band detected catalogues. These objects must be located in the field-of-view covered by the $U_{\rm k}$ band, since the LBGs are selected from this band. They must also have NIR data in order to have the best constraints on the photometric redshift. All known LBGs, BBGs, ISAAC ghost images, spurious detections and double entries were removed and the appropriate magnitude cuts were applied. A sample of 96 objects was obtained of which 93 were detected in $R$ band and 3 in either of the $K_{\rm s}$ bands. 

\begin{figure}
\resizebox{\hsize}{!}{\includegraphics{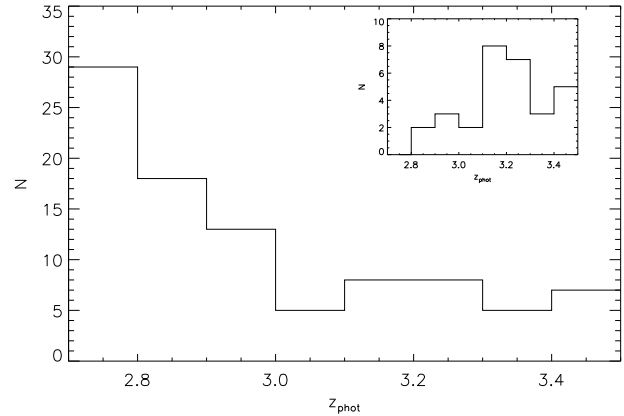}}
\caption{\label{fig:restz} Redshift distributions of `non-classified' objects detected in $R$ band that have $2.7 < z_{\rm phot} < 3.5$ and $R \le 26$. The inset shows the subset of objectts that satisfy the redshift, $V_{\rm v}-R$ and magnitude criteria for LBG selection but are undetected in $U_{\rm k}$.}
\end{figure}

Figure~\ref{fig:restz} shows the photometric redshift distribution of the $R$ band detected galaxies. The distribution differs significantly from the $z_{\rm phot}$ distribution of LBGs as there is a large number of objects located at the lower edge of the redshift range. 
 
The majority of galaxies with $3.0 < z_{\rm phot} < 3.3$ is undetected in the $U_{\rm k}$ band and was assigned lower limits for their $U_{\rm k}-V_{\rm v}$ colours. These $U_{\rm k}-V_{\rm v}$ lower limits are generally too blue for these objects to be classified as LBGs, but deeper $U$ band data might lead to redder $U_{\rm k}-V_{\rm v}$ colours, shifting them into the LBG region. The inset in Fig.~\ref{fig:restz} shows the $z_{\rm phot}$ distribution of these possible LBGs. A peak at the redshift of the protocluster is seen. This peak accounts for a large fraction of the unclassified objects with $3.0 < z_{\rm phot} < 3.3$. There are 55 potential LBGs (pLBGs) that have lower limits on $U_{\rm k}-V_{\rm v}$ that are too blue to be classified as LBGs. The LBG selection therefore misses up to 50 per cent of $z\sim 3$ galaxies.

In the $K_{\rm s}$ band only three additional objects are found to be missing from the BBG sample. The same redshift range yields 15 unique BBGs. Based on this we estimate that the BBG selection yields $\sim80$ per cent of all $z\sim3$ $K_{\rm s}$ detected galaxies.

\begin{figure*}
\includegraphics[width=120mm]{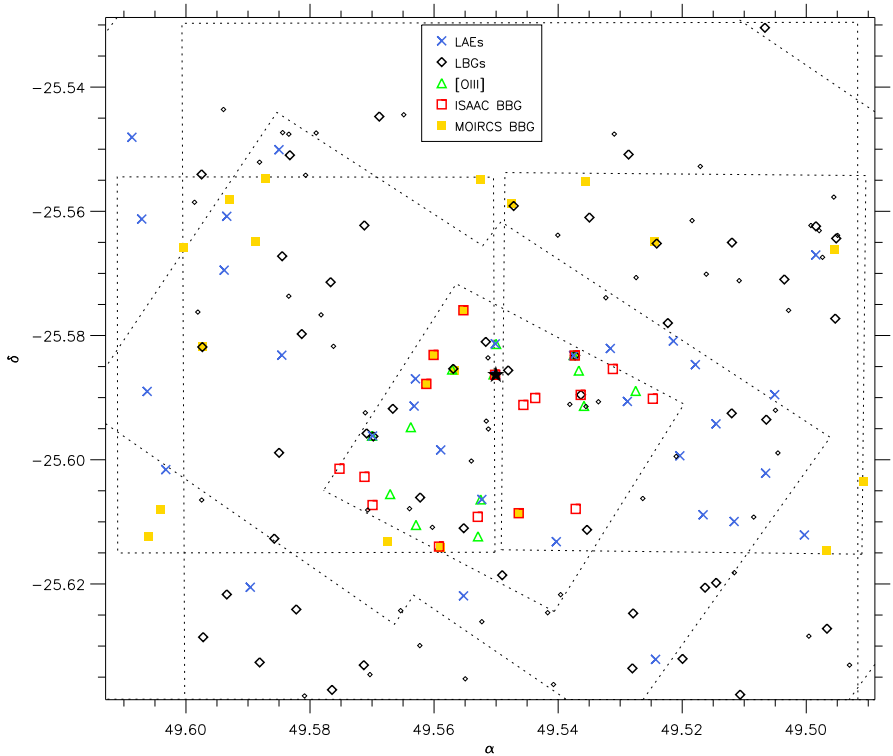}
\caption{\label{fig:allobs} Sky coordinates of the various populations. Large and small diamonds indicate the LBG and pLBGs, respectively. Blue crosses denote LAEs, green triangles [O{\sc iii}] emitters and the squares denote the BBGs. A distinction is made between ISAAC and MOIRCS detected BBGs, with the former indicated by open red squares and the latter by filled yellow squares. The dotted lines indicate the outlines of the fields covered by the various instruments. The FORS data ($V_{\rm f}I$) covers the entire figure and also includes all VIMOS bands ($U_{\rm v}BV_{\rm v}R$). Subsequently, the IRAC data covers the majority of the field with the exception of some of the corners. Going from larger to smaller fields we then have the $U_{\rm k}$ field, the field covered by ACS, the two MOIRCS fields and finally, slightly below centre the area covered by ISAAC.}
\end{figure*}

The positions of the objects in the final samples as well as the fields covered by the various instruments used are shown in Fig.~\ref{fig:allobs}.

\section{SED fitting}  \label{sec:fitting}

Do the properties of the protocluster galaxies differ from those of field galaxies at the same redshift? To compare the protocluster galaxies to field galaxies we determine the galaxy properties by fitting the individual galaxy SEDs with population synthesis models, and then compare the properties to equivalent samples detected in the field. 

The fitting of the galaxy SEDs was done using the FAST code described in \citet{kriek2009}. Due to its versatility and speed the use of the FAST code allows us to fit a large range of models and model parameters. Both the BC03 and the updated CB07 models (Charlot~\&~Bruzual, private communication) were used with a Salpeter IMF and $Z=$\Zsun. The inclusion of an improved treatment of the TP-AGB phase in the CB07 models results in a more reliable stellar mass estimate for populations older than 100~Myr.

The free parameters in the fitting routine are the age, mass, SFH and the extinction by dust. We considered exponentially declining SFHs with decay times, $\tau$, ranging from 10~Myr to 10~Gyr with steps of 0.1~dex. The inclusion of SFHs with values of $\tau$ much larger than the age of the Universe for our adopted cosmology allowed us to mimic a constant SFH. The age grid that we considered ranges from log(age/yr)=7 to 9.3 with age steps of 0.1~dex, where log(age/yr)=9.3 equals the age of the Universe at $z\sim3.13$. The redshifts were fixed to the spectroscopic or photometric redshifts, depending on which was available. As the EAZY code is specialized in determining redshifts and is more efficient in use, we have chosen to use the EAZY redshifts rather than including it as a free parameter during the SED fitting. This choice has little effect on the main results for the majority of the objects.

The effect of internal dust extinction the model SEDs was taken into account using the \citet{calzetti2000} extinction law for values of $A_{\rm V}$ ranging from 0 to 3 with steps of 0.1. The attenuation blueward of the Ly$\alpha$ line due to the IGM was included using the prescription of \citet{madau1995}.

\section{Results} \label{sec:res}

\subsection{Number densities of galaxy populations} \label{sec:ndens}

\subsubsection{Lyman Break Galaxy candidates} \label{sec:nlbgs}

To determine whether there is an overdensity of LBGs in the 0316 field due to the presence of the protocluster, the number density of LBGs in the 0316 field is compared to the $\sim0.3\square^{\circ}$ MUSYC ECDF-S blank field (see also Sect.~\ref{sec:samplbg}). Photometry for the control field was obtained as described in Sect.~\ref{sec:photom}.

The GOODS-S field, which is part of the ECDF-S field, has been shown to be underdense in DRGs \citep{vandokkum2006}. Comparison of the number density of DRGs and LBGs in 8 ECDF-S  subfields reveals that the variation in LBG number density is lower than for DRGs. Furthermore, the central subfield, which has the lowest DRG number density, has a relatively high number density of LBGs. We conclude that the ECDF-S is unlikely to be as underdense in LBGs as the GOODS-S field is in DRGs. 

For the MUSYC ECDF-S data an LBG colour criterion was devised that is equivalent to the criterion used for the 0316 field:
\begin{eqnarray} \label{eq:mcrit}
& U-V \ge 1.85, & \nonumber \\
& V-R \le 0.62, & \\
& U-V \ge 4.0\times (V-R)+2.51, & \nonumber \\
& R \le 26. & \nonumber
\end{eqnarray}
Applying this to the MUSYC data yields a total number of 694 LBG candidates.

\begin{figure}
\resizebox{\hsize}{!}{\includegraphics{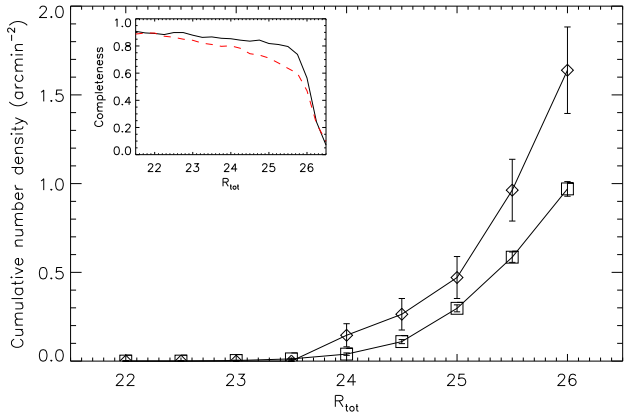}}
\caption{\label{fig:odens2} completeness-corrected cumulative number density of objects that satisfy the colour selection criterion as function of total $R$ band magnitude. Diamonds denote values for the 0316 field, whereas the squares indicate the results found for the MUSYC ECDF-S control field.  Uncertainties are based on Poisson statistics. An excess of LBGs in the 0316 field is apparent. The inset shows the completeness of the fields as a function of magnitude with the full and dashed line denoting the 0316 and ECDF-S data, respectively.}
\end{figure}

Figure~\ref{fig:odens2} shows the completeness-corrected cumulative number density of LBGs for both the 0316 and the ECDF-S control field. Note that no information on the redshifts of the objects is used to select these LBG samples. There is an excess in the 0316 field across the entire magnitude range. The 0316 field is a factor of $1.6\pm 0.3$ denser than the ECDF-S field for galaxies with $R\le 25.5$, with the 1$\sigma$ uncertainty based on Poisson statistics. Defining the galaxy surface overdensity as $\delta_{\rm g} = n_{\rm 0316}/n_{\rm ECDF-S}-1$, we find $\delta_{\rm g}=0.6\pm 0.3$ in the 0316 field. This overdensity is not the result of a difference in general number counts, as the area-normalized completeness-corrected number counts of the two fields differ by less than 1.5 per cent. 

\begin{figure}
\resizebox{\hsize}{!}{\includegraphics{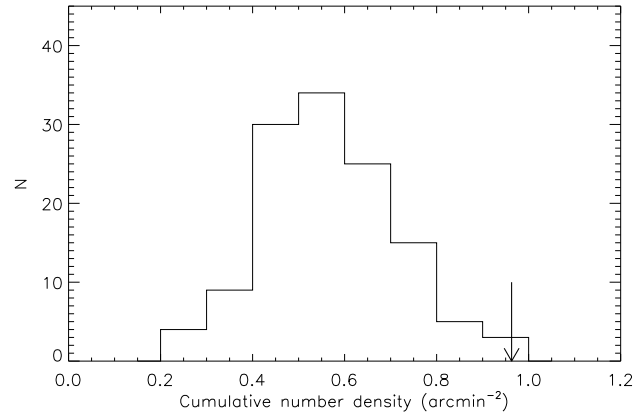}}
\caption{\label{fig:odenshist} Distribution of LBG number densities with $R\le25.5$ in the ECDF-S for 125 randomly chosen Keck-sized ($40.1\square$\arcmin) fields. The arrow indicates the value found for the 0316 field. One in 125 ECDF-S subfields is found to host a larger number density than found in the 0316 field.}
\end{figure}

The significance of the LBG overdensity in 0316 is quantified by considering field-to-field variations of LBGs on the scale of the 0316 field.  The number density of LBGs in 125 subfields of the ECDF-S (each the same size as the 0316 field) is measured, and the resulting distribution displayed in Fig.~\ref{fig:odenshist}; the arrow indicates the number density of LBGs in the 0316 field. Only 1 in 125 fields has a number density that exceeds the 0316 density, thus the significance of the LBG overdensity in the 0316 field is at the 3$\sigma$ level with respect to field-to-field variations.

The surface overdensity of LBGs in the 0316 field is a lower limit to the volume overdensity of the protocluster in this field. The LBG selection criterion selects galaxies in a relatively large redshift range, so the LBG overdensity within the protocluster is larger than 0.6 (see Sect.~\ref{sec:disc}).

The outer VIMOS quadrants are used to assess the LBG number density at greater distances from the RG. There is no $U_{\rm k}$ coverage for these outer fields, so the criterion given in Eq.~\ref{eq:crit} cannot be used. Instead, a similar criterion that uses the $U_{\rm v}$ filter is applied. The number density in the outlying fields is approximately 10 per cent smaller than the central field. Therefore these outlying fields are overdense with respect to the ECDF-S, but to a lesser degree than the central 0316 field. This indicates that the protocluster extends beyond the central field, out to at least 15\arcmin~($\sim7$~Mpc). This is larger than most estimates of protocluster sizes which typically find sizes of 2--5~Mpc \citep{intema2006,venemans2007}, but often size estimates are limited by the field size. Nevertheless, the larger size may be a consequence of the possible superstructure in the field, as hinted at by the redshift distribution of the [O {\sc iii}] emitters.
 
\subsubsection{Balmer Break Galaxy candidates} \label{sec:ndrgs}

The study of K07 found that the 0316 field is overdense with respect to the GOODS-S field, with approximately 1.5--2 times more BBGs in the 0316 field. However, \citet{vandokkum2006} has pointed out that the GOODS-S field is underdense, containing only 60 per cent of the number of objects found in the larger MUSYC survey \citep{gawiser2006,quadri2007}. To ascertain whether there is an overdensity of BBGs in the 0316 field, we compare to the publically available MUSYC data described in \citet{quadri2007}. 

The four MUSYC fields cover a total area of $\sim$400$\square\arcmin$. Using these fields as control fields will yield better statistics compared to the GOODS-S, as well as decrease the influence of cosmic variance. The control fields are denoted as HDFS1, HDFS2, 1030 and 1255, respectively. Catalogues were constructed from the four images following the procedure used for the 0316 field. Applying the BBG criterion and a $K_{\rm s}\le 23$ cut yields 123, 116, 197 and 118 BBGs for the HDFS1, HDFS2, 1030 and 1255 fields, respectively, or 554 BBGs in total. 

\begin{figure}
\resizebox{\hsize}{!}{\includegraphics{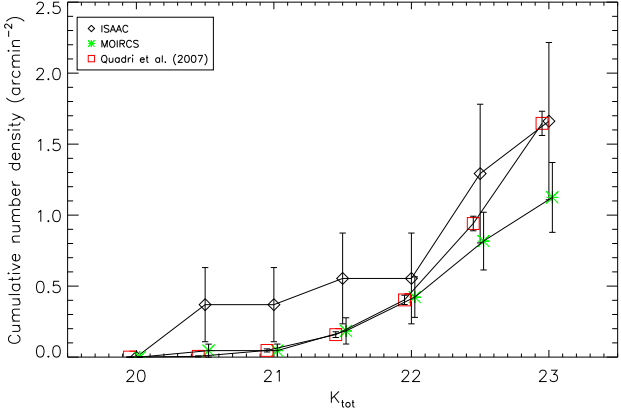}}
\caption{\label{fig:odensdrg}Completeness-corrected cumulative number density as function of $K_{\rm s}$ band magnitude for both populations of BBGs. ISAAC and MOIRCS detected BBGs are denoted by diamonds and asterisks, respectively. Uncertainties are obtained using Poisson statistics. The squares denote the mean cumulative number density of the four individual MUSYC fields. Neither of the 0316 BBG populations shows evidence for an overdensity in the 0316 field.}
\end{figure}

Fig.~\ref{fig:odensdrg} compares the completeness-corrected cumulative number density of BBGs in the 0316 field to the MUSYC control field. The small field-of-view ISAAC data suggests that the 0316 field has an excess of bright BBGs ($K_{\rm s}\le 21.5$) close to the radio galaxy, but the sample is small so the number statistics are poor. At fainter magnitudes there is no overdensity. Since the MOIRCS field shows no indication of an excess at bright magnitudes, we conclude that there is no evidence for an overdensity of BBGs in the 0316 field. This is in contradiction with the conclusion of K07.

The lack of an overdensity seems to contradict the presence of a peak in the photometric redshift distribution at the protocluster's redshift (Fig.~\ref{fig:zphotdrg}). Nine ISAAC selected BBGs lie in the range $2.7 < z_{\rm phot} < 3.5$. Only four of these have $K_{\rm i} \le 23$, so more than half of the BBGs at the protocluster's redshift are too faint for inclusion in Fig.~\ref{fig:odensdrg}. It is possible most of the protocluster BBGs are faint, and would only be revealed as an overdensity at faint magnitudes, however the control fields are not deep enough to warrant a proper comparison.

\subsection{Properties of protocluster galaxy candidates} \label{sec:physprop}

Below we present the results of the SED fitting for all protocluster populations. The LAEs are treated separately in Sect.~\ref{sec:laes}, because they are generally very faint across the entire wavelength range. This results in poorly constrained SED fits and therefore a different approach was taken for this population. 

\subsubsection{Mass} \label{sec:mass}

\begin{figure}
\resizebox{\hsize}{!}{\includegraphics{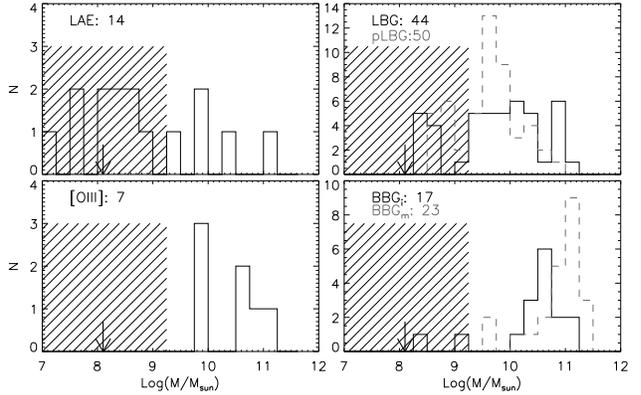}}
\caption{\label{fig:popmass}Stellar mass distributions for each protocluster population. Objects that are strongly affected by confusion in the IRAC images, or galaxies with $M \le 10^7$~\Msun~are not included. The total number of objects in each population is shown in the upper left corner of each panel with the same colour coding as the respective histograms. The arrow indicates the best-fitting mass that is found for the LAE stack (Sect.~\ref{sec:laes}). The hatched region indicates where mass estimates become highly uncertain due to faintness in the restframe optical.}
\end{figure}

The mass distributions of all protocluster candidate populations are shown in Fig.~\ref{fig:popmass}. The estimate of the galaxy's stellar mass is dominated by the flux in the IRAC bands, so all objects that are strongly contaminated by neighbouring sources have been excluded from the samples. Table~\ref{table2} lists the median mass values for each population. 

The upper right panel of Fig.~\ref{fig:popmass} shows that most LBGs have masses between 10$^9$ and 10$^{11}$~\Msun~and the median stellar mass is a few times 10$^{9}$~\Msun. This mass increases to $\sim2\times$10$^{10}$~\Msun~when limiting the sample to LBGs with $R\le 25.5$. The median mass of the pLBG population is approximately a factor 2 lower than the LBGs, and a KS test shows that the mass distributions differ at the $2\sigma$ level. Thus the $z\sim3$ galaxies that are not included in the LBG sample due to insufficient depth of the $U$ band are generally less massive than the LBG galaxies.

A large fraction of the [O{\sc iii}] emitters is detected in the [3.6] and [4.5] bands, whilst a few are also detected in the [5.8] and [8.0] bands. Six objects (the RG and source IDs 2, 6, 10, 11 and 12 in M08) are strongly contaminated by nearby bright objects. The remaining 7 objects are all detected in the [3.6] and [4.5] bands. This implies that the [O{\sc iii}] emitters have significant stellar masses. The median stellar mass of the [O{\sc iii}] emitters is larger than the median LBG mass, however, this median mass is based on a sample of only 7 objects.

[O{\sc iii}] emitters are selected as high equivalent width objects in the observed $K_{\rm s}$ band, thus they are selected to have low $K_{\rm s}$ band magnitudes relative to the [O{\sc iii}] narrowband\footnote[15]{At low redshifts the [O{\sc iii}] line is often found to be an indicator of AGN activity. Therefore part of the IRAC flux may be caused by heated dust close to an AGN.}. Hence it is surprising that such a selection criterion identifies massive galaxies.

Not all [O{\sc iii}] emitters have been spectroscopically confirmed, so it is possible the sample contains low redshift interlopers. The prime suspects for interlopers would be H$\alpha$ emitters at $z\sim 2.15$.  Three out of  the 7 [O{\sc iii}] emitters with reliable mass estimates are not spectroscopically confirmed and two of these have $z_{\rm phot}\sim2.17$. Moreover, all spectroscopically confirmed [O{\sc iii}] emitters have $z_{\rm phot} > 3$. Thus it is possible that these two objects are interlopers. Removing these objects from the sample leaves three galaxies with $M=$6--8$\times 10^{9}$~\Msun~and two galaxies with $\sim10^{11}$~\Msun.

The BBGs have masses of a few times 10$^{10}$~\Msun~with some exceeding $10^{11}$~\Msun. The BBGs comprise the population with the highest masses of the protocluster candidate populations. Since the BBGs are $K_{\rm s}$ band selected it is expected that they will have high stellar masses. The ISAAC-selected BBGs are on average less massive than the MOIRCS-selected BBGs, but this is only due to the larger depth of the ISAAC data. When the same magnitude cut is applied to both datasets the mass of the ISAAC-selected BBGs agrees with that of the MOIRCS-selected BBGs.

\subsubsection{Age and extinction} \label{sec:age}

The SED fitting procedure results in degeneracies between the best-fit age, extinction and SFH. Because of the large uncertainties in the best-fit ages of the LBGs, we do not discuss them further in this work.  The ages of the BBGs are easier to constrain, because they were selected to have strong Balmer breaks. The median age at which the BBGs started forming stars is $\sim$1~Gyr ago (see Table~\ref{table2}). 

\subsubsection{UV slope} \label{sec:uvbeta}

Assuming the restframe UV continuum has the form $f_{\lambda}=C \lambda^{\beta}$, we calculated the UV slope $\beta$ of the candidate protocluster members. Either the $R$ and $I$ or $r_{625}$ and $I_{814}$ bands were used, depending on whether ACS coverage is available. At $z\sim 3$ these bands correspond to restframe 1500~\AA~and 2000~\AA, respectively. The UV slope was calculated as
\begin{equation}
\beta=\frac{\log{f_{\lambda,\rm R}} - \log{f_{\lambda,\rm I}}}{\log{\lambda_{\rm eff,R}} - \log{\lambda_{\rm eff,I}}}
\end{equation}
with $\lambda_{\rm eff,R/I}$ the effective wavelength of the $R$ or $I$ filters, respectively. The uncertainty on $\beta$ was obtained by varying the flux by an amount drawn randomly from the normal photometric error distribution.  

The median $\beta$ for the LBG population $\beta$ is -1.52, but the bright subset ($R\le 25.5$) has a redder UV slope of -1.38. This reddening of $\beta$ with increasing brightness was also found by \citet{bouwens2009}. An unobscured, young population has a UV slope ranging from $-2$ to $-2.5$. Thus the LBGs are slightly attenuated by dust. The pLBGs are on average bluer than the LBGs, indicating there is less dust in these objects.

The [O{\sc iii}] emitters have similar UV colours to the LBGs. Whereas the BBGs have significantly redder colours than the LBGs so they are either more heavily obscured or older than the LBGs. 

\subsubsection{Star formation rate} \label{sec:sfrres}

The star formation rates (SFRs) of the candidate protocluster galaxies were estimated from the rest-frame 1500~\AA~luminosity using the relation from \citet{kennicutt1998}
\begin{equation}
SFR_{\rm UV} [M_{\odot} yr^{-1}] = \frac{L_{\rm 1500}}{8\times 10^{27}}.
\end{equation}
This relation is valid for a Salpeter IMF and does not take the presence of dust into account. For simplicity we assume that all objects are located at the redshift of the protocluster. Furthermore, this relation is only valid for starbursts that have reached an equilibrium between the number of stars forming and evolving off of the main sequence. The SFRs of extremely young starburst galaxies that have not yet reached this equilibrium will be  underestimated.  

\begin{figure}
\resizebox{\hsize}{!}{\includegraphics{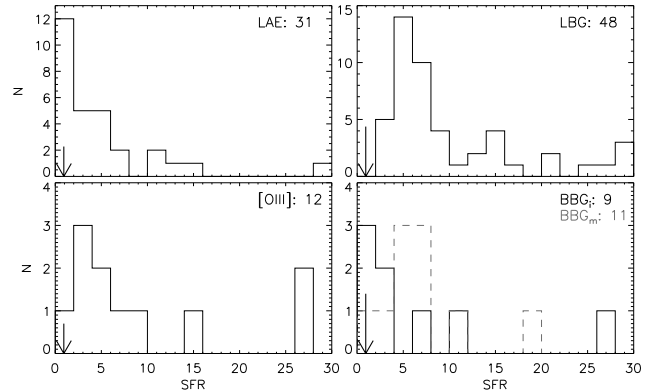}}
\caption{\label{fig:popsfr}Extinction uncorrected star formation rates for the various populations obtained from the restframe 1500~\AA~luminosity. The arrow indicates the SFR obtained for the LAE stack. For simplicity, we have assumed that all LBGs and BBGs with $2.7 < z < 3.5$ are located at the protocluster's redshift.}
\end{figure}

The SFRs of the candidate protocluster galaxies are shown in Fig.~\ref{fig:popsfr}. Most galaxies have SFRs less than 10~\Msun~yr$^{-1}$, irrespective of galaxy population, with a small fraction of outliers having SFRs up to 40~\Msun~yr$^{-1}$. Both the median dust-uncorrected and dust-corrected SFRs for each population are listed in Table~\ref{table2}. The extinction was derived from the UV slopes using $E(B-V)=(\beta-\beta_{\rm 0})/8.067$ \citep{meurer1995} and $A(1500)=4.39E(B-V)$ \citep{calzetti2000}. Here $\beta_{\rm 0}$ is the UV slope for an unattenuated ionizing population of stars, assumed to be $-2.5$. This method of estimating SFRs gives good agreement with SFRs derived from radio observations for galaxies at $z\sim2$ \citep{pannella2009}.

The LBGs have a relatively small amount of dust extinction with a median $A(1500)=0.53$ ($A_{\rm V}=0.21$). \citet{shapley2001} found generally larger dust extinction with a median value of $A_{\rm V}\sim0.6$. We revisit this difference in a more thorough comparison in Sect.~\ref{sec:discproplbg}. No significant difference was found between the uncorrected SFRs of the LBGs and pLBGs, but the dust-corrected SFRs of the pLBGs are systematically lower, consistent with the UV slopes of the pLBGs being systematically bluer.

The slope of the UV continuum can become redder due to both aging of the stellar populations, and attenuation by dust. Thus it is ambiguous to use the UV slope as an indicator of extinction for BBGs. Therefore the amount of extinction was measured from both the UV slopes and the value for $A(1500)$ derived from the SED fitting. Both values of the dust-corrected SFR are listed in Table~\ref{table2}.

The estimated SFRs of the MOIRCS BBGs are systematically larger than those of the ISAAC BBGs. However, the MOIRCS data are shallower than the ISAAC data and therefore include objects with systematically larger mass. The observed difference in SFRs between MOIRCS and ISAAC objects is consistent with the increase of SFR with increasing mass at $z<1$ \citep{noeske2007} and at $z\sim2$ \citep{daddi2007}.

\begin{table*}
\begin{center}
\caption{\label{table2} Upper part: median values of various properties for the samples discussed in this work. Uncertainties here are mean absolute deviations on the median divided by the square root of the number of objects. Lower part: Properties for the stacked and bright, individual LAEs.}
\begin{tabular}{l|c|c|c|c|c|c|l|}
\hline
Object & $\log(M)_{\rm CB07}$ & $\log(M)_{\rm BC03}$ & SFR (\Msun~yr$^{-1}$) & SFR$_{\rm corr}$ (\Msun~yr$^{-1}$) & log(Age)$_{\rm CB07}$ &$\beta$\\
\hline
\hline
LBG & 9.92 & 10.1 & 7.0 & 11.3 & -$^{\rm a}$ & -1.52$\pm0.07$ \\
LBG$_{R<25.5}$ & 10.16 & 10.29 & 11.4 & 23.1 & -$^{\rm a}$ & -1.38$\pm 0.09$ \\
pLBG & 9.65 & 9.79 & 5.7 & 9.1 & -$^{\rm a}$ & -1.72$\pm0.08$ \\
pLBG$_{R<25.5}$ & 9.79 & 9.95 & 8.7 & 13.9 & -$^{\rm a}$ & -1.54$\pm 0.15$ \\
O{\sc iii}$^{\rm b}$ & 10.51 & 10.62 & 6.3 & 9.5 & -$^{\rm a}$ & -1.50$\pm 0.13$ \\
BBG$_{\rm I}$ & 10.55 & 10.77 & 2.2$^{\rm c}$ & 5.1/6.9$^{\rm c,d}$ & 9 & -1.12$\pm0.21$$^{\rm c}$ \\
BBG$_{\rm M}$ & 11.0 & 11.15 & 6.6$^{\rm c}$  & 15.0/6.8$^{\rm c,d}$ & 8.9 & -0.91$\pm0.19$$^{\rm c}$ \\
\hline
LAE$_{\rm stack}$ & 8.1$^{+0.42}_{-0.3}$ & 8.1$^{+0.53}_{-0.32}$ & 0.9 & 1.0 & 7.0$^{+1.0}$ & -2.40$\pm 0.43$ \\
LAE$_{\rm 1518}$ & 9.97$^{+0.11}_{-0.19}$ & 10.07$^{+0.14}_{-0.19}$  & 15.7 & 20.4 & 8.8$^{+0.1}_{-0.2}$ & -1.97$\pm 0.32$\\
LAE$_{\rm 1867}$ & 11.08$^{+0.05}_{-0.12}$ & 11.22$^{+0.08}_{-0.11}$  & 40.2 & 64.2 & 8.8$^{+0.1}_{-0.3}$ & -1.56$\pm 0.73$ \\
LAE$_{\rm 3101}$ & 10.48$^{+0.22}_{-0.68}$ & 10.74$^{+0.09}_{-0.94}$ & 11.3 & 27.6 & 8.8$^{+0.5}_{-1.7}$ & -0.7$\pm 1.8$\\
\hline
\hline
\multicolumn{7}{l}{$^{\rm a}$ Individual ages are unconstrained}\\
\multicolumn{7}{l}{$^{\rm b}$ Values for the SFRs and $\beta$ taken from M08}\\
\multicolumn{7}{l}{$^{\rm c}$ Only objects with $2.7 < z < 3.5$ are taken into consideration}\\
\multicolumn{7}{l}{$^{\rm d}$ First value calculated using the UV slope and second value calculated using the SED fit results}\\
\hline
\end{tabular}
\end{center}
\end{table*}

\subsubsection{Ly$\alpha$ emitters} \label{sec:laes}

In the sample of LAEs, 27 are not detected in any IRAC band, indicating that the LAEs have small stellar masses. Figure~\ref{fig:popmass} shows that most LAEs have $M<10^{9}$~\Msun. The properties of these IRAC-undetected LAEs were investigated by mean stacking 16 LAEs that lie within the field-of-view of the $U_{\rm v}BVRI$ and MOIRCS datasets and are not contaminated by bright neighbours in the IRAC data. 

\begin{figure}
\resizebox{\hsize}{!}{\includegraphics{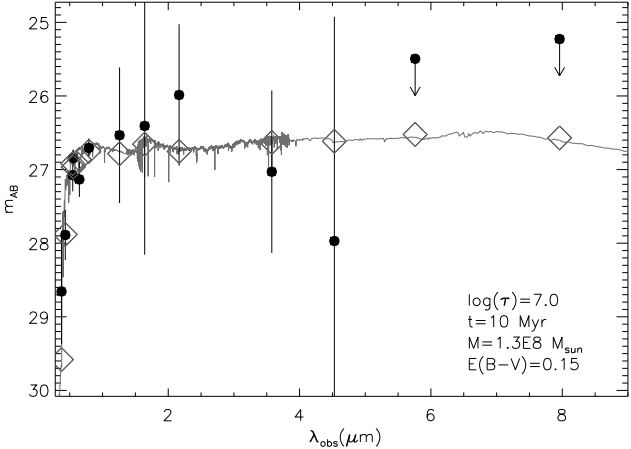}}
\caption{\label{fig:stack}Photometry of a stack of 16 LAEs and the corresponding best-fitting SED. The diamonds indicate the photometry as computed from the best fit SED. Best-fitting parameters are listed in the bottom right corner. 1$\sigma$ upper limits are used for [5.8] and [8.0] as the stack yields negative flux in these bands. The best-fitting model is very young and has a low mass, but a significant amount of dust extinction. Due to the large uncertainties on the JHK and IRAC photometry a model SED without dust would lead to an equally acceptable fit.}
\end{figure}

The photometry and the best-fitting SED for the stacked data are shown in Fig.~\ref{fig:stack}. The stack reveals robust detections in the $BVRI$ bands, but not in bands redward of $I$. The derived stellar mass is approximately $10^8$~\Msun and this is marked in Fig.~\ref{fig:popmass} with an arrow. It is much smaller than the average LBG stellar mass. The best-fit age is 10~Myr, which is the lowest age allowed by the SED fitting process. The amount of dust extinction is unconstrained from the fit to the SED, but the UV slope determined from the $r_{\rm 625}$ and $I_{\rm 814}$ bands is $\beta=-2.4$. Thus the average protocluster LAE contains little dust. These results are in agreement with previous studies that show that LAEs are in general young objects experiencing their first burst of star formation \citep[e.g.][]{hu1996,malhotra2002,tapken2004}.

Only 4 of 32 LAEs have robust detections in [3.6] and [4.5], whilst 1 LAE has a marginal detection. The four objects with significant IRAC detections include the RG.  However, since the RG is located close on the sky to a foreground galaxy, the photometry in all bands is compromised. It is therefore not included in further analysis \citep[but see also][who estimate its mass to be 1.6$\times 10^{11}$~\Msun~based on a Kroupa IMF]{seymour2007}. The remaining 4 LAEs are discussed below using the IDs designated in V05.

\begin{figure*}
\resizebox{\hsize}{!}{\includegraphics{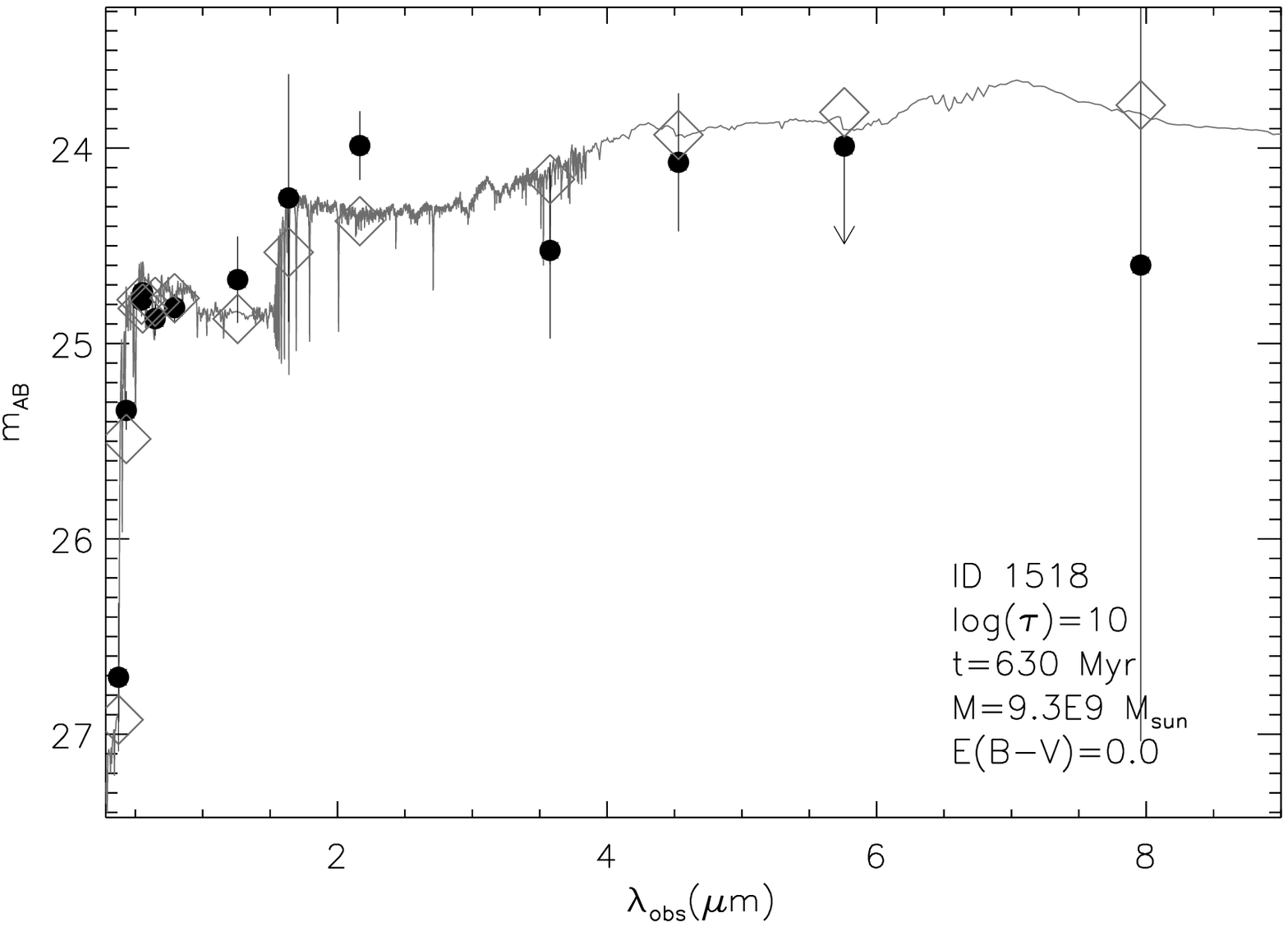}\includegraphics{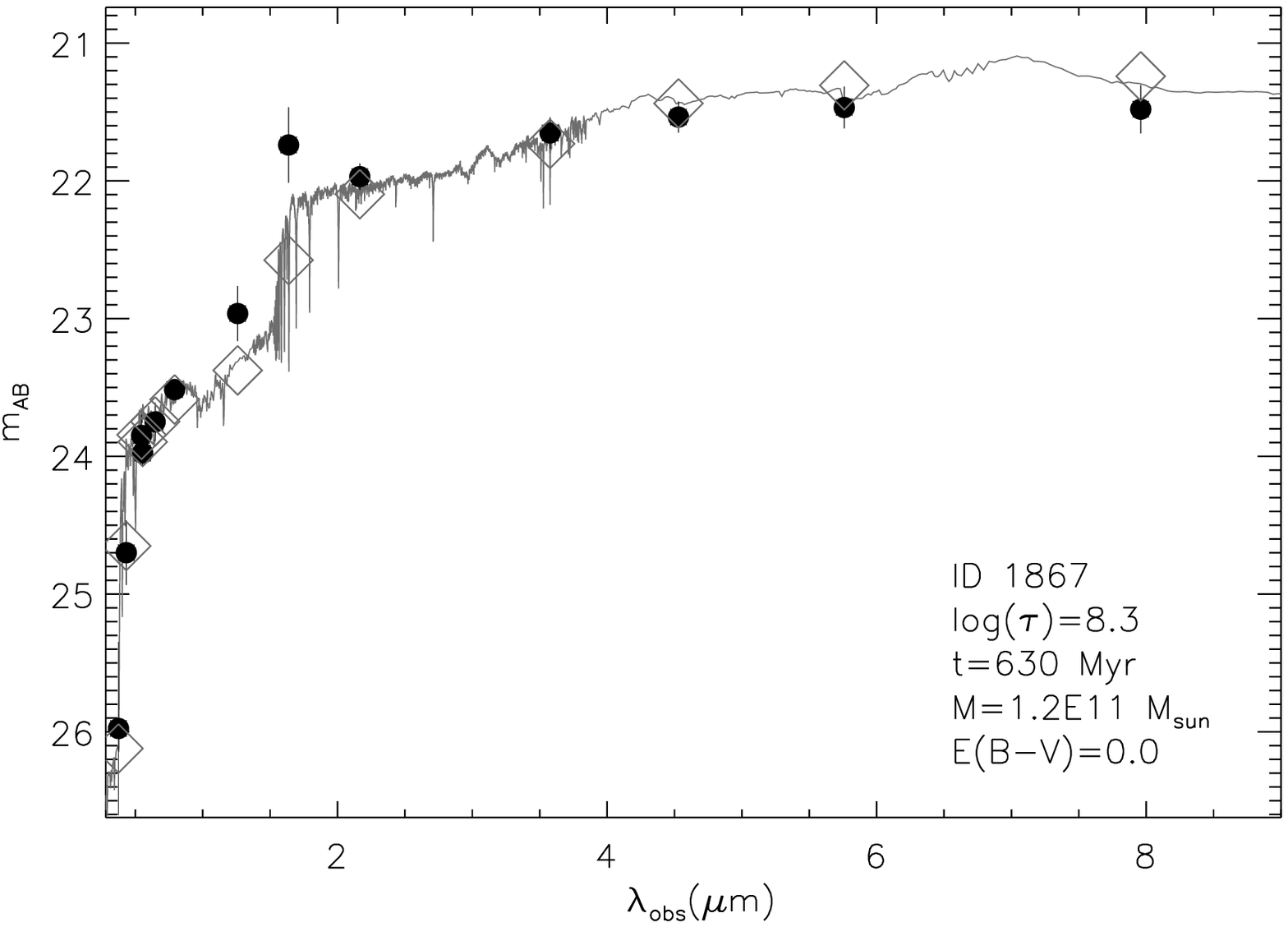}}
\resizebox{\hsize}{!}{\includegraphics{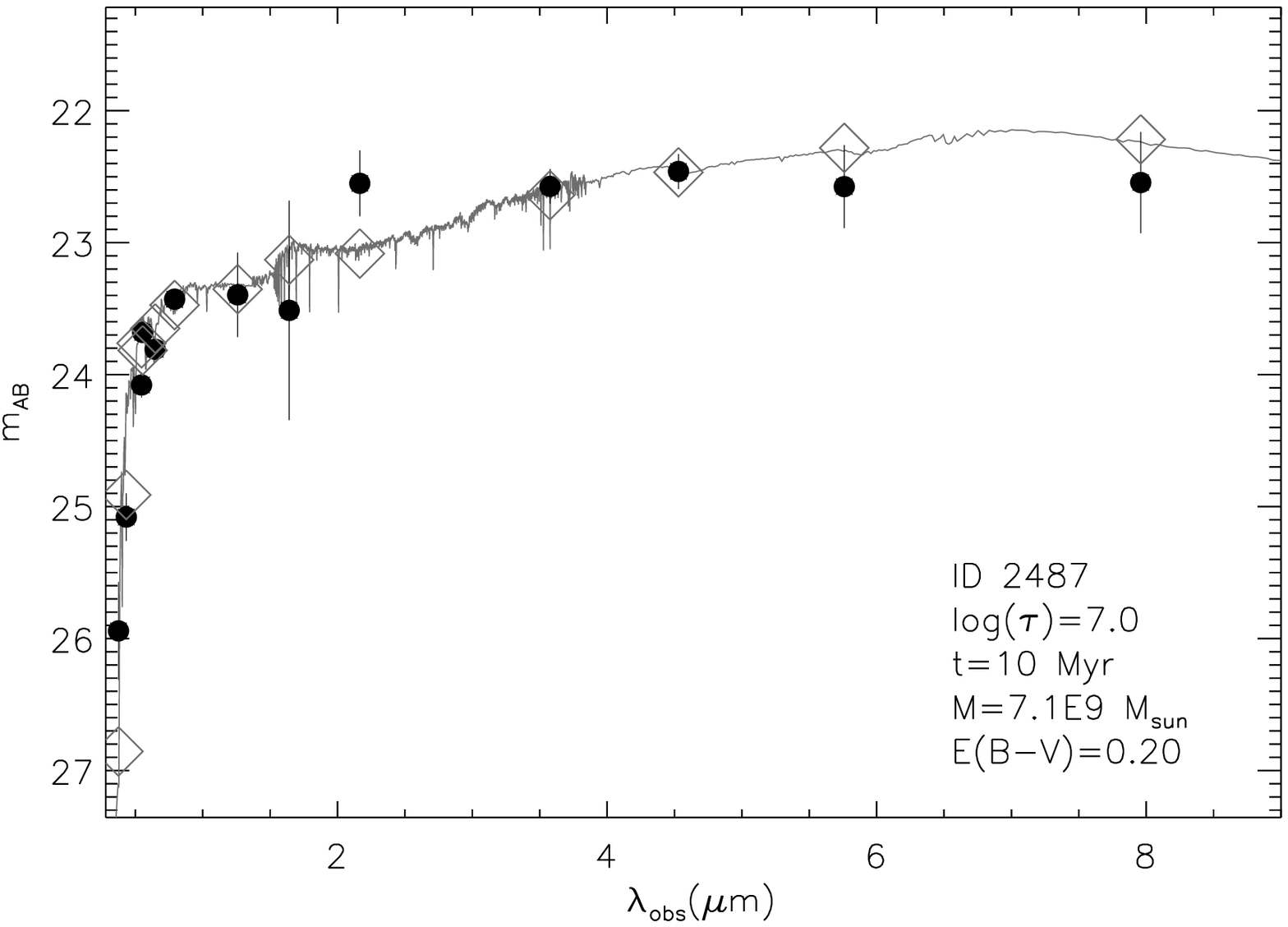}\includegraphics{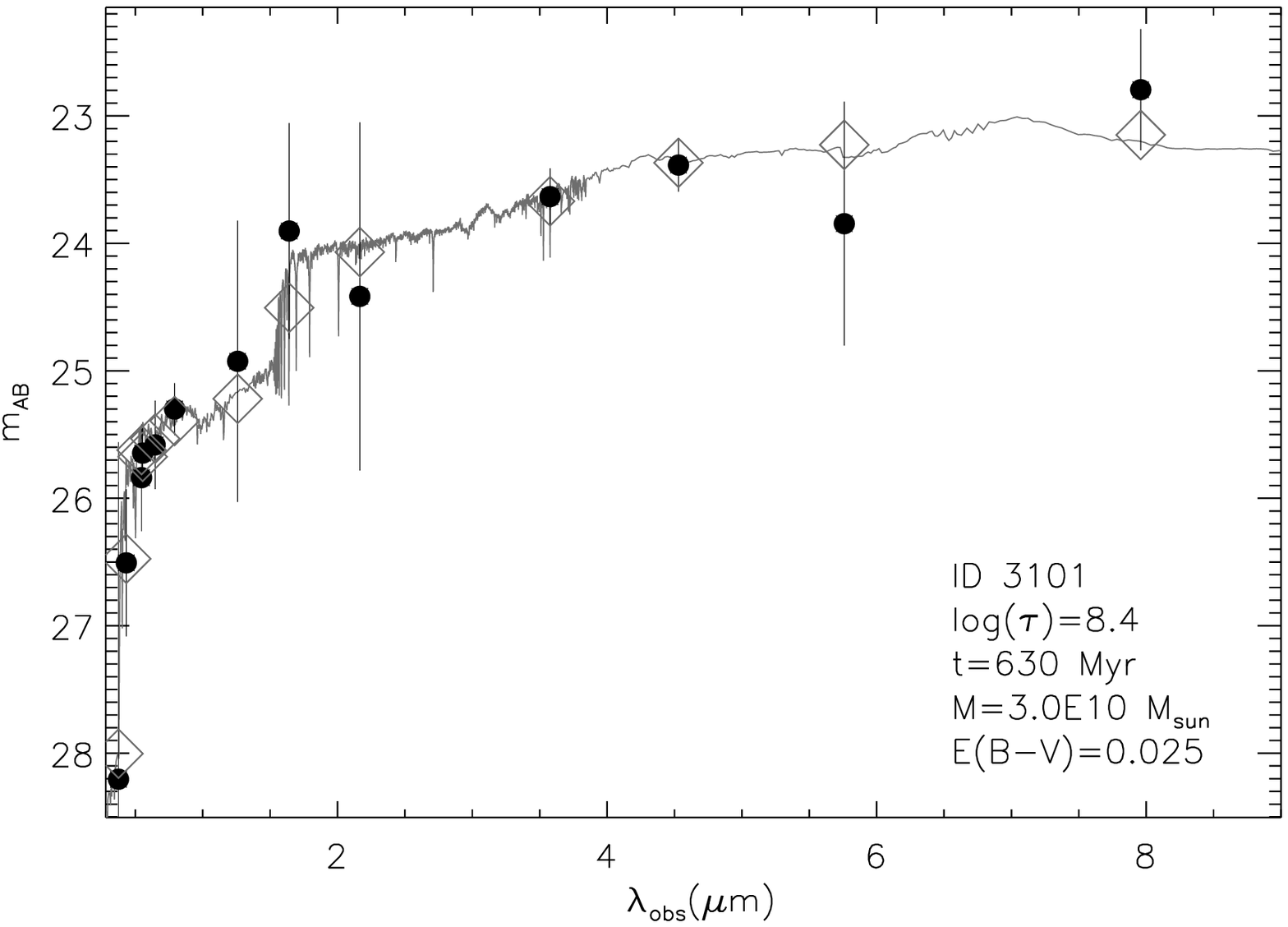}}
\caption{\label{fig:4fits} Photometry and SED fits for the four LAEs that were detected in the [3.6] and [4.5] bands. Best-fitting parameters are shown in the bottom right corner of each panel. Diamonds denote magnitudes as calculated from the best-fitting SED. ID numbers correspond to the ones used in V05. The first two objects fall inside the limited ISAAC field-of-view and are shown with the ISAAC $J$ and $K_{\rm s}$ data for the restframe optical. For the remaining two objects no ISAAC data was available and the shallower MOIRCS $JHK_{\rm s}$ data is used. Three out of four SED fits return ages of 500~Myr and higher and all masses are approximately larger than $5\times10^{9}$. Non-detections are replaced with $1\sigma$ upper limits.}
\end{figure*}

{\bf 1518:} The detection of this object in the [3.6] and [4.5] bands is marginal. However, it has significant detections in both $J$ and $K_{\rm i}$ (6$\sigma$ and 7.5$\sigma$, respectively). The photometry and the best-fit SED are shown in the upper left panel of Fig.~\ref{fig:4fits}. The galaxy is best fit by a model with continuous SFR and an age of 600~Myr. No dust attenuation is required. The best-fit stellar mass is $\sim10^{10}$~\Msun, making it almost two orders of magnitude more massive than the average LAE.

{\bf 1867:} Excluding the RG, this object is the second brightest LAE in the sample. It has a clear IRAC detection in all four bands and is also bright in the restframe UV/optical continuum. The photometric redshift of this object is 3.15, which is consistent with the spectroscopic redshift of 3.134. The galaxy is of similar age to object \#1518 with a best fit age of 600~Myr and $\tau\sim200$~Myr. It contains no dust and it is very massive, having a stellar mass of approximately 10$^{11}$~\Msun. This galaxy is also selected as an LBG and a BBG. It has a clumpy morphology as seen in ACS data (V05, Venemans et al. in prep.) and is located 50\arcsec\ (or 350~kpc in projection) from the RG. 

{\bf 2487:} This object hosts an AGN, as evidenced by the broadness of the Ly$\alpha$ line and the presence of a broad CIV emission line \citep[V05,][]{lefevre1996}. Since the SED fitting does not account for an AGN contribution, the best fit parameters are not valid and are not discussed further. 

{\bf 3101:} This LAE has similar properties to \#1867, but it is less massive by a factor 3. Because of the absence of ISAAC data, the best-fit properties are poorly constrained. 

An overview of the properties including 1$\sigma$ uncertainty intervals can be found in Table~ \ref{table2}. Note that the two objects for which only MOIRCS NIR data is available (\#2487 and \#3101) have significantly larger uncertainties on the best-fitting age. This reflects the aforementioned need for deep NIR observations in order to constrain the age.

\subsection{Dependance of galaxy properties on location within the protocluster} \label{sec:spatial}

An important question in studying the history of cluster evolution is whether there are systematic variations in the properties of the protocluster members that depend on their location within the protocluster, e.g. distance from the radio galaxy.

\begin{figure}
\resizebox{\hsize}{!}{\includegraphics{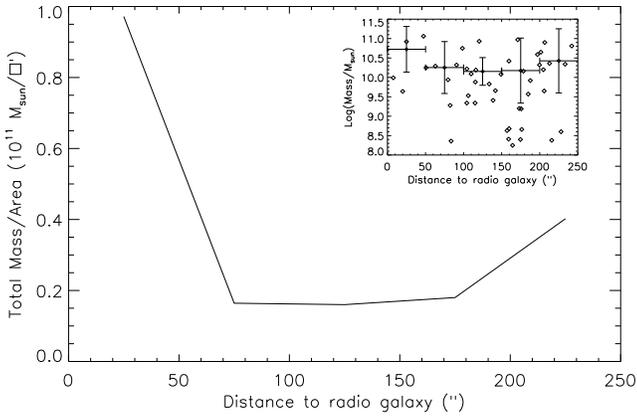}}
\caption{\label{fig:massrad} Stellar mass of all LBGs as function of distance to the RG. Masses of individual objects within annuli of 50\arcsec\ width are summed and normalized to the area of the annuli. The inset shows the individual values as function of distance to the RG and mean values for each of the annuli. The individual values show a weak trend, but the integrated mass shows a sharp decrease as the distance to the RG increases.}
\end{figure}

In Fig.~\ref{fig:massrad} the stellar mass of the LBGs is plotted as a function of distance from the RG. The main panel of Fig.~\ref{fig:massrad} shows the area-normalized integrated mass $M_{\rm int}$ within annuli of 50\arcsec, whilst the inset shows the individual masses and the mean mass per annulus. 
Two massive LBGs near the RG result in a high $M_{\rm int}$ in the innermost annulus. $M_{\rm int}$ then sharply decreases out to 150\arcsec, but slightly increases at greater distances.

The inset in Fig.~\ref{fig:massrad} shows a large scatter in the masses of the LBGs at all radii, which is expected with such a large number of interlopers. This large scatter in combination with the small sample size could also cause the upturn at distances greater than 150\arcsec. 

The two most massive LBGs are located near the RG: one is a confirmed LAE protocluster member, whereas the other is an [O{\sc iii}] emitter with $z_{\rm phot}=3.23$. Both galaxies have stellar masses of approximately $10^{11}$~\Msun~making them almost as massive as the RG itself (2.5--3$\times10^{11}$~\Msun~ for a Salpeter IMF). This is different from what is found for radio-selected protoclusters at $z=$4--5, where the RG is approximately an order of magnitude more massive than the brightest of the surrounding LBGs \citep{overzier2009}. 

\begin{figure}
\resizebox{\hsize}{!}{\includegraphics{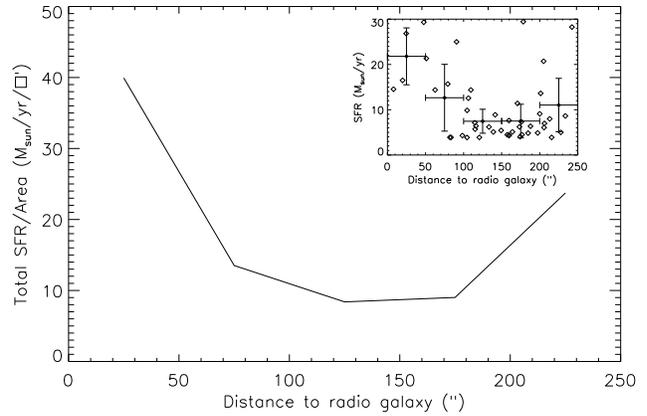}}
\caption{\label{fig:sfrrad}Total area-normalized SFRs for annuli of 50\arcsec\ width. SFRs are dust-uncorrected. The inset shows the individual values and the mean values for bins of 50\arcsec. Vertical error bars are mean absolute deviations. The seemingly lower limit to the SFR present in the inset is due to the magnitude cut at $R=26$ used to select LBGs. Objects close to the RG are forming stars more rapidly.}
\end{figure}

Figure~\ref{fig:sfrrad} shows the area-normalized integrated SFR in annuli of 50\arcsec. A similar trend is observed for the SFR as for $M_{\rm int}$: galaxies close to the RG have larger SFRs, although again the scatter is large. At small distances ($<100$\arcsec) all objects are forming stars at rates of $>$15~\Msun~yr$^{-1}$, whereas at larger distances the majority of the objects are forming stars at $\sim$5~\Msun~yr$^{-1}$. That galaxies with the highest SFR are located in the centre of the protocluster is in stark contrast with the situation in galaxy clusters at $z<1$ where the SFR drops towards the cluster core \citep[e.g.][]{balogh1998,balogh1999,hashimoto1998,postman2001,lewis2002}.

\begin{figure*}
\resizebox{\hsize}{!}{\includegraphics{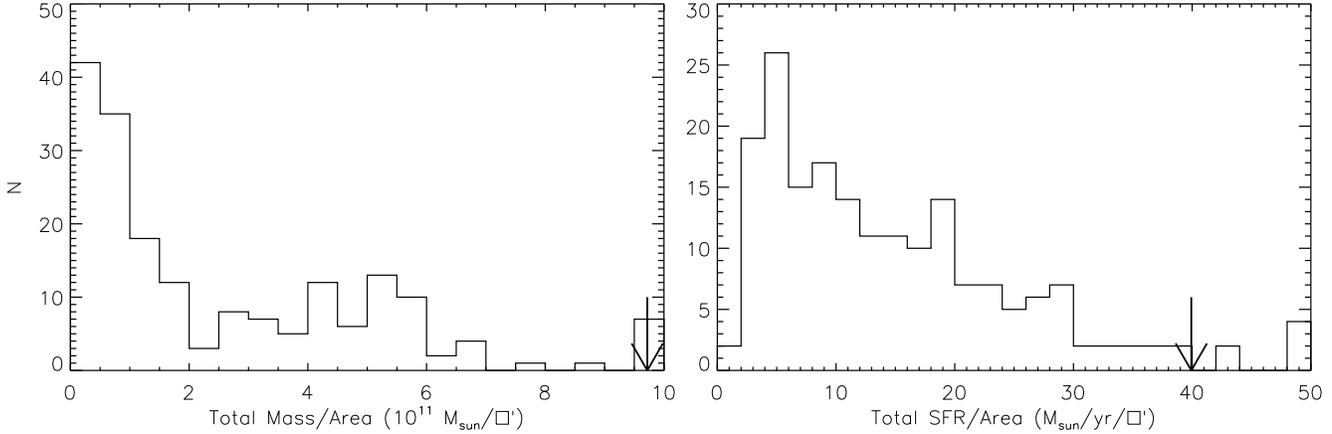}}
\caption{\label{fig:checkrad}Left panel: Distribution of total area-normalized mass within a 50\arcsec~radius for 200 random locations across the 0316 field. Right panel: Same as left panel but for SFR. Arrows indicate the total area-normalized mass and SFR in a 50\arcsec~annulus when the RG is taken as the centre of the protocluster.}
\end{figure*}

The above figures show that the RG is surrounded by some of the most massive and active galaxies in the protocluster. How unique is the position of the RG in the protocluster? The above procedure was repeated for 200 random locations in the 0316 field, and the resulting mass and SFR distributions for the innermost annulus are shown in Fig.~\ref{fig:checkrad}. There are few locations that yield similar or larger $M_{\rm int}$ or total area-normalized SFR, and these are all located close to the RG. Thus the immediate surroundings of the RG are unique within the 0316 field. The radio galaxy likely lies in the core of the forming protocluster and the protocluster core galaxies are affected by their environment. 

Note that there are several effects that will weaken such observed spatial trends. First, since the photometric redshifts alone cannot establish protocluster membership, there will be contamination by interlopers in such an analysis. The observed LBG surface overdensity of 0.6 implies that slightly less than 2 out of every 3 LBGs is an interloper. Secondly, many protocluster systems show filamentary structures \citep[e.g.][]{ouchi2005,matsuda2005,overzier2008}, hence the assumption of a circular symmetry used above will also weaken the signal. Thirdly, the RG itself is not included in this analysis, and its large mass ($>10^{11}$~\Msun) and SFR (110~\Msun~yr$^{-1}$ without dust correction) will lead to an even stronger correlation if included. Thus the radial dependence is likely to be stronger than observed.

We conclude from the observed radial trends in stellar mass and SFR that galaxy evolution is strongly influenced by the protocluster environment, with larger SFRs and stellar masses occurring in the protocluster core.

\section{Discussion} \label{sec:disc}

\subsection{Surface and volume overdensities} \label{sec:odensdisc}

\subsubsection{Lyman Break galaxies}

The measured LBG surface overdensity of the 0316 field is $0.6\pm 0.3$. Because the uncertainties of the photometric redshifts ($\sigma_{z}/ (1+z)$) range between 0.02 and 0.35, the derived surface density will be diluted by the fore- and background objects that are in the sample as well. The surface overdensity of 0.6$\pm 0.3$ that is found is thus a lower limit to the volume overdensity of the LBGs. How strong must the volume overdensity of LBGs be to produce a surface overdensity of 0.6?

V05 showed that the FWHM of LAE redshifts in the 0316 protocluster is $\Delta z\sim 0.025$. LAEs are in general significantly less massive than LBGs so they may relax within the protocluster potential more easily. We assumed that the protocluster spans a line-of-sight distance of $\Delta z=0.04$ (as found in the study of \citet{steidel1998} for LBGs in the SSA22 protocluster at $z=3.09$). In the field-of-view covered by the $U_{\rm k}$ band, the protocluster spans a volume of $\sim5.6\times 10^{3}$~Mpc$^{3}$. 

The LBG selection criterion selects galaxies in the redshift interval $2.9 < z < 3.4$, as shown in the inset in Fig.~\ref{fig:colcol}. Although the spread in the $z_{\rm phot}$ distribution exceeds this interval,  this can be attributed to the uncertainties in the $z_{\rm phot}$ estimates. There are 448 LBGs with $R<25.5$ in the ECDF-S and these span a volume of 1.8$\times 10^{6}$~Mpc$^{3}$. This implies that there should be approximately 16--17 'field' LBGs in the 0316 field. 

To obtain a surface overdensity of $\delta_{\rm g}=0.6$ an additional 10--11 LBGs are needed in the protocluster volume. The protocluster volume can host approximately 1.3 'field' LBGs. Hence the protocluster is a factor $\sim8\pm 4$ denser than the field and the volume overdensity $\delta_{\rm g,V}$ is $7\pm4$, with the 1$\sigma$ uncertainties calculated by propagation of the uncertainty on the surface overdensity. This is larger than the overdensities observed in the LAEs and [O{\sc iii}] emitters, but this volume overdensity is little more than an order of magnitude estimate due to the many uncertainties in the assumptions that were made.

Furthermore, based on the presence of three [O{\sc iii}] emitters at $z=3.1$, M08 speculated the 0316 protocluster is part of a larger 60 comoving Mpc superstructure. This superstructure would extend beyond the redshift range probed by the Ly$\alpha$ NB filter. If the protocluster is part of a larger superstructure, the protocluster volume assumed above would result in an overestimate of the LBG overdensity.

The mass of the protocluster was computed using the comoving volume $V$ occupied by the protocluster, the mean density of the Universe $\bar{\rho}$ and the matter overdensity $\delta_{\rm m}$ which relates to the galaxy volume overdensity $\delta_{\rm g,V}$ as $\delta_{\rm m}=\delta_{\rm g,V}/b$ with $b$ being the bias parameter.  Using $V=5.6\times10^{3}$~Mpc, $\bar{\rho}=1.9\times10^{10}$~\Msun~Mpc$^{-3}$ and $b=$1--5 \citep[e.g.,][]{adelberger1998,giavalisco1998,ouchi2004,lee2006} the protocluster mass is $M=$2--12$\times10^{14}$~\Msun. The range was determined by the combination of the 1$\sigma$ uncertainty and the range in the bias value. This protocluster mass is consistent with V05 and is similar to the mass of the Virgo cluster. Since it was shown in Sect.~\ref{sec:nlbgs} that the overdensity is spread over multiple VIMOS quadrants, the protocluster volume is likely underestimated. Therefore, the cluster mass should be considered a lower limit.

\subsubsection{Balmer Break galaxies}

We found no surface overdensity of BBGs in the 0316 field based on the number densities. But would we expect to detect a surface overdensity in the BBGs assuming the protocluster has the same volume overdensity of BBGs as it has of LBGs? Or will the overdensity be smoothed out due to the large redshift range that is probed by the BBG selection criterion?

The 554 BBGs detected in the four MUSYC fields are assumed to lie between $2 < z < 3.5$ and are homogeneously spread across this redshift interval. Thus the volume spanned by this population is 2.2$\times10^{6}$~Mpc$^3$. This implies that there should be 7--8 'field' BBGs with $K_{\rm s}<23$ in the ISAAC field-of-view. The protocluster occupies 755~Mpc$^3$ (within the ISAAC field-of-view), and this volume contains on average 0.2 'field' BBG. So if the BBGs are overdense by the same factor as the LBGs (volume overdensity of 7), the protocluster will host $\sim1.4$ BBG. This results in a surface number density in the ISAAC field that is 1.2 times as high as the field density and thus a surface overdensity $\delta_{\rm g}=0.2$. This is consistent with the 1$\sigma$ uncertainties on the cumulative number densities at $K_{\rm s}=23$ as shown in Fig.~\ref{fig:odensdrg}. We conclude that a volume overdensity of 7 cannot be ruled out due to the small sample size.

\begin{figure}
\resizebox{\hsize}{!}{\includegraphics{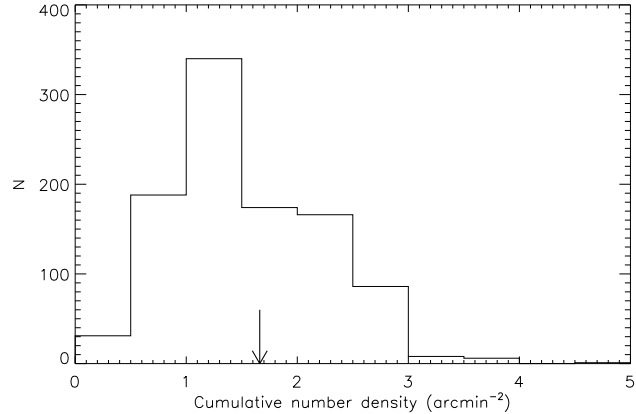}}
\caption{\label{fig:drgfluc}Distribution of BBG number densities with $K_{\rm s} \le 23$ in the UDS field \citep{williams2009} for 1000 randomly chosen ISAAC-sized (5.4$\square$\arcmin~fields. The arrow indicates the respective number density found for the 0316 field. To obtain a 3$\sigma$ overdensity signal with respect to the field-to-field variations a number density of at least 3.5/$\square$\arcmin is needed.}
\end{figure}

It is doubtful whether systematic searches for BBG overdensities at $z\sim3$ have any chance of succeeding considering the range in redshift that is probed. To evaluate this the amplitude of field-to-field variations for an ISAAC-sized field is assessed. A similar method as described in Sect.~\ref{sec:nlbgs} is employed. As reference the UDS catalogue from \citet{williams2009} is used. This field is complete down to $K_{\rm s}=23$ and covers a total of $\sim0.8\square^{\circ}$. The BBG number density was evaluated in 1000 ISAAC-sized fields placed randomly across the UDS field. The resulting distribution is shown in Fig.~\ref{fig:drgfluc}. The arrow indicates the BBG density found in the 0316 field. It is clear that this result is not significant with respect to the field-to-field variations. In fact, a number density of $\sim3.5$ per $\square$\arcmin~is needed to obtain a surface overdensity that is significant at the $3\sigma$ level. This would imply a surface overdensity $\delta_{\rm g}\sim1.2$, equivalent to a volume overdensity of $\sim45$. Such a strong overdensity is highly unlikely to exist over a large field-of-view. We conclude that selection criteria that span a considerable redshift range are poorly suited for the detection of overdensity signals around HzRGs.

\subsection{Field and cluster ensemble properties} \label{sec:discprop}

Does the evolution of galaxies in crowded environments in the Early Universe, such as protoclusters, differ significantly from galaxy evolution in the field? At $z=2.3$, \citet{steidel2005} found that both ages and masses of galaxies in a protocluster are $\sim2$ times larger than similarly selected galaxies in the field. A similar study at $z=$4--5 has shown no significant difference between protocluster and field galaxies \citep{overzier2009}. This study can give a better insight in the situation at $z \sim 3$. 

We compare galaxy masses, SFRs and UV slopes for the various protocluster populations with those of equivalent populations in the field. There is no sample of field [O{\sc iii}] emitters in the literature, hence this population will not be discussed. We also refrain from comparing the 0316 BBGs to field BBGs as there is no strong evidence for an overdensity in the 0316 field. The SED fit results given here have been obtained using BC03 models with a Salpeter IMF to facilitate comparison with previous studies.

\subsubsection{Ly$\alpha$ emitters}

Studies of LAEs at $z\sim 3$ find stellar masses ranging between 10$^8$ and 10$^9$~\Msun~\citep{gawiser2006b,nilsson2007,lai2008} which is fully consistent with the results presented in this study. 

We also presented SEDs for a small population of massive LAEs that do not fit the picture of galaxies that experience their first burst of star formation. This subpopulation of massive LAEs is found frequently in recent studies. \citet{finkelstein2009} found that two objects in their sample of 14 LAEs at $z=4.4$ were best fit by a model SED with an age of 800~Myr and a mass of 6.5$\times 10^9$~\Msun. This fraction of 2 out of 14 is similar to the fraction of massive objects found in this study. \citet{lai2008} find a fraction of massive LAEs of approximately 30 per cent, but they quote a $2\sigma$ detection limit of 25.2 for [3.6]. This is approximately half a magnitude deeper than the 0316 data. \citet{pentericci2009} investigated Ly$\alpha$-emitting LBGs, finding a small fraction of objects that are significantly older than the average LAE. In summary, there is no evidence that the LAEs in the protocluster are significantly different from the field population.

\subsubsection{Lyman break galaxies} \label{sec:discproplbg}

Taking all LBGs with $R \le 25.5$, the median mass of the 0316 LBGs is $\sim1.6$ times more massive than that of 'field' LBGs, which have a median mass of $1.2\times 10^{10}$~\Msun~\citep{shapley2001}. In addition, LBGs close to the RG are more massive than LBGs at larger radii, indicating protocluster environment promotes the increased stellar mass of its members. However, the BC03 population synthesis models were not available at the time of the study of \citet{shapley2001}, so a fully consistent comparison can not be made.

\citet{lehmer2009} find that LBGs suspected to be in the protocluster SSA22 at $z=3.09$ have on average $H$ band luminosities that are 1.2--1.8 times larger than LBGs that are unrelated to the protocluster. Since the $H$ band can be interpreted as a proxy for stellar mass at $z\sim3$ this also indicates that the masses of the protocluster LBGs are larger by the same factor. This factor of 1.2--1.8 is similar to the difference in median masses found in this study.

The 0316 LBGs have dust-uncorrected SFRs of $\sim$7--11~\Msun~yr$^{-1}$ (depending on whether the whole sample is considered or only the objects that have $R \le 25.5$). The $\mathcal{R}$ data presented in \citet{steidel2003} yields a median dust-uncorrected SFR for field LBGs of $\sim 10$~\Msun~yr$^{-1}$. This is consistent with the median dust-uncorrected SFR found for the 0316 LBGs.  If the SFRs are corrected for dust extinction, the SFRs increase on average by  a factor of 2. This small increase in SFR indicates the presence of small amounts of dust in the LBGs.  \citet{shapley2001} used SED fitting to find that field LBGs have a median extinction of $E(B-V)=0.155$ ($A_{\rm V}=0.62$). This is larger than found in the 0316 protocluster as half of the 0316 LBGs are best fit by $A_{\rm V}\le0.1$. However, $A_{\rm V}$ is poorly constrained, with 31 out of 48 LBGs having $1\sigma$ uncertainties that are consistent with $A_{\rm V}=0.6$. We can therefore not conclude whether the 0316 LBGs are less dusty than field LBGs.

\citet{meurer1999} composed a sample of $z\sim3$ LBGs from the HDF V2.0 catalog \citep{williams1996} using a $UVI$ filter combination. This filter set is similar to the set used in this study. They find a median  UV slope ($\beta$) for field LBGs of approximately -1.6, similar to 0316 LBGs. We thus find no evidence for a difference in the UV slopes between field and cluster LBGs at $z\sim3$. \citet{shapley2003} find significantly redder slopes for field LBGs, but the filter set used in their work differs significantly from this study, so it does not offer a fair comparison.

We find that the 0316 LBGs have stellar masses, SFRs and UV slopes that are consistent with those of field LBGs. We thus conclude that at $z\sim3$ the ensemble properties of all possible protocluster galaxies is not significantly different from what is found in a low-density environment.

\subsection{The future of MRC~0316--257} \label{sec:future}

We have found that the 0316 field contains an overdensity of star forming galaxies. How will the protocluster LBGs evolve and at what epoch will a quiescent population emerge?

Using a simple `back of the envelope' calculation we estimate how the protocluster will appear at later times. The peak of the cosmic SFH lies below $z\sim3$ \citep[e.g.][]{steidel1999} and therefore new LBGs will emerge at later times. These `new' objects are not taken into account and we only evolve  the LBGs that are currently identified. The sample is therefore static and no new objects are introduced.

The SED fitting yields both a best-fitting age and SFH, so it is possible to evolve the LBGs according to this information. We find, however, that both these quantities are poorly constrained. Hence, we make the following assumptions. Since the distribution of SFHs peaks at approximately  $\tau=10^{8}$~yr, we assume that the typical LBG has an SFH of 100--500~Myr. An age of 50--500~Myr is assumed, in accordance with earlier studies that state 100~Myr to be the typical age for LBGs at $z \sim 3$ \citep{shapley2001}. Finally, the initial SFR is assumed to be 40--60~\Msun~yr$^{-1}$. These SFRs are to be interpreted as being intrinsic SFRs. They are based on the results from \citet{shapley2001} and as shown in Sect.~\ref{sec:discprop} are consistent with the results for the 0316 LBGs. For simplicity the influence of dust extinction will not be taken into account in this scenario.

If a quiescent galaxy is defined as having an SFR $\le 2$~\Msun~yr$^{-1}$ we find that the majority of a population of LBGs will have reached quiescence after a period of approximately 1~Gyr. This corresponds to $z\sim2$. Only the youngest LBGs having $\tau=500$~Myr will last until $z\sim1.5$ before they reach quiescence. Therefore, the 0316 protocluster should harbour a substantial `dead' population after 1~Gyr of evolution.

These predictions can be compared to observations of protocluster structures at lower redshifts. One of the best studied protoclusters is the overdense region surrounding the $z=2.16$ radio galaxy PKS~1138--262 (1138) \citep{miley2006}. This radio galaxy has a intricate, clumpy morphology and seems to be in an advanced stage of merging with several smaller galaxies. It is surrounded by a diffuse UV halo which is evidence for in-situ star formation outside the main radio galaxy \citep{hatch2008}. Furthermore, the field has been shown to harbour overdensities in both Ly$\alpha$ and H$\alpha$ emitting galaxies, extremely red objects (EROs), X-ray emitters and sub-mm bright galaxies \citep{pentericci2000,kurk2004a,kurk2004b,stevens2003,croft2005,zirm2008,kodama2007}. \citet{kurk2004a} has also shown that the EROs and H$\alpha$ emitting galaxies are located closer to the radio galaxy than the bluer star forming galaxies. The 1138 protocluster is therefore a prime example of a $z\sim0$ galaxy cluster progenitor.

\citet{zirm2008} have shown that the 1138 system contains red galaxies, some of which might be quiescent. However, the red sequence is still in the stages of formation and has not been properly established yet. This is in agreement with studies by \citet{blakeslee2006} and \citet{vandokkum2007} that state that the formation redshift of the red sequence lies in the range $z=$2--2.5. This roughly agrees with our toy model for the 0316 protocluster. A significant number of red galaxies should be in place as early as $z=2.5$. This is earlier than what most studies predict but we are limited by the simplicity of this calculation. If the SFH is episodic and burst-like the time it takes for an LBG to become quiescent increases, so we expect the red sequence will form at a later time than estimated by this toy model.

We predict that the population of galaxies currently residing in the 0316 protocluster will resemble the 1138 protocluster after 1~Gyr of evolution and 0316, like the 1138 protocluster, will evolve into a rich galaxy cluster at $z<1$.

\section{Summary and conclusions} \label{sec:conc}

We have presented a comprehensive study of several galaxy populations in the protocluster associated with MRC 0316-257 at $z=3.13$, using restframe FUV to optical images. In addition to studying the LAEs and [O{\sc iii}] emitters found in previous studies, we identify samples of LBG and BBG candidates whose properties provide additional evidence for the presence of a protocluster.

\begin{enumerate}
\item{
The cumulative number density of the LBG candidates in the 0316 field is a factor of 1.6$\pm$0.3 larger than for comparable non-protocluster fields at similar redshifts, indicating a surface overdensity of 0.6$\pm$0.3 for LBGs in the protocluster. The surface overdensity is significant at the $3\sigma$ level with respect to the field-to-field variations. This measured LBG surface overdensity strengthens the conclusion of V05 that a protocluster surrounds MRC 0316-257. Using estimates of the protocluster size, the measured surface overdensity gives an LBG volume density that is $8\pm$4 larger than that of the field. Such an overdensity implies a minimum mass for the protocluster of 2--12$\times10^{14}$~\Msun.}

\item{
The redshift distribution of the BBG candidates shows a peak at the redshift of the radio galaxy. This is further evidence that there is an overdensity of galaxies and a protocluster around MRC 0316-257. There is no significant surface overdensity of BBGs, but this is not surprising due to the small sample size. A volume overdensity of BBGs comparable with the measured volume density of LBGs could not have been detected. Selection criteria that probe large redshift ranges are poorly suited for such overdensity studies.

We are unable to reproduce the results obtained by K07 who found that the 0316 field is a factor 1.5--2 denser in BBGs than blank fields. We attribute this to the choice of control field. According to \citet{vandokkum2006} the GOODS-S field, used by K07, is underdense in $2<z<3$ red galaxies by a factor of $\sim$2. In this study four control fields are used with a total area exceeding 400$\square\arcmin$, reducing the effect of cosmic variance on the control fields.
} 

\item{
The masses and ages of candidate protocluster galaxies were determined using SED fitting. LAEs generally have faint continuum emission at 3.6 $\mu$m and 4.5~$\mu$m and relatively blue UV slopes, indicating that they contain little dust. Typical stellar masses of the LAEs are $\sim 10^{8}$~\Msun. Four of the LAEs have larger IRAC fluxes and stellar masses in excess of 10$^{10}$~\Msun. This supports results in previous studies that there is also a non-negligible number of older evolved LAEs
}

\item{
The median mass determined for the protocluster LBGs is a factor of two larger than that for the field LBGs. Although this is not significant, considering the uncertainties of SED fitting, it agrees with the results of \citet{lehmer2009}, who found that protocluster LBGs are more luminous at $H$ band and therefore more massive than field galaxies. No significant difference between cluster and field LBGs is found for SFRs and UV-slopes.
}

\item{
The most massive and intensely star forming galaxies are located primarily near to the radio galaxy indicating that proximity to the radio galaxy influences galaxy evolution. No other high mass and high SFR regions are found in the 0316 field, indicating that the radio galaxy is near the centre of the protocluster. The trend in SFR is radically different from what is observed in local galaxy clusters where the SFR decreases systematically towards the cluster centre.
}
\end{enumerate}

We conclude that the protocluster surrounding MRC 0316-257 is in a relatively early stage of formation. The blue LBG population will likely evolve into a population of passive, red galaxies over a timescale of 1~Gyr, forming a structure similar to the well-studied protocluster surrounding MRC 1138-262 at $z = 2.2$

Spectroscopic follow-up of the LBG and BBG candidates is needed to refine their redshifts and establish which individual objects are indeed located within the protocluster. Furthermore, population studies of a large sample of radio-selected protoclusters over a range of redshift are needed to obtain a more comprehensive view of how these structures form and evolve. Such studies would clarify whether the protocluster around MRC~0316--257 is representative of large-scale overdensities in the early Universe.

\section*{acknowledgements}

We wish to thank the anonymous referee for all the useful suggestions that have improved this paper significantly. This research has been based on observations made with the VLT at ESO Paranal, programs 072.A-0284(A), 077.A-0310(A,B), 078.A-0002(A,B) and 167.A-0409(A,B).  Also based on observations made with the NASA/ESA $HST$, obtained at the Space Telescope Science Institute (STScI). STScI is operated by Association of Universities for Research in Astronomy, Inc., under NASA contract NAS~5-26555. This study is also based on data collected at Subaru Telescope, which is operated by the National Astronomical Observatory of Japan. The W.M. Keck Observatory is a scientific partnership between the University of California and the California Institute of Technology, made possible by a generous gift of the W.M. Keck Foundation. The authors wish to recognize and acknowledge the significant cultural role and reverence that the summit of Mauna Kea has always within the indigenous Hawaiian community; we are fortunate to have the opportunity to conduct observations from this mountain. EK acknowledges funding from Netherlands Organization for Scientific Research (NWO). NAH and GK acknowledge funding from the Royal Netherlands Academy of Arts and Sciences (KNAW). The work by SAS at LLNL was performed under the auspices of the U.S. Department of Energy under Contract No. W-7405-ENG-48 and in part under Contract DE-AC52-07NA27344. This work is based [in part] on observations made with the Spitzer Space Telescope, which is operated by the Jet Propulsion Laboratory, California Institute of Technology under a contract with NASA. Support for this work was provided by NASA through an award issued by JPL/Caltech. SC acknowledges support for radio galaxy studies at UC Merced, including the work reported here, with the Hubble Space Telescope and Spitzer Space Telescope via NASA grants HST \#10127, SST \#1264353, SST \#1265551 and SST \#1279182.

\end{document}